\documentclass[11pt]{article}
\usepackage[T2A]{fontenc}
\usepackage{amsfonts}
\usepackage[sumlimits]{amsmath}
\usepackage{amsthm}
\usepackage{amssymb}
\usepackage{setspace}
\usepackage{graphicx}

\newcommand{\p}{\partial}
\newcommand{\be}{\begin{eqnarray}}
\newcommand{\ee}{\end{eqnarray}}
\newcommand{\de}{\delta}
\newcommand{\eps}{\varepsilon}

\newtheorem{propose}{Proposition}
\newcommand{\prop}[1]{\begin{propose} \  #1 \end{propose}}

\newtheorem{definn}{Definition}

\newcommand{\sls}{\nonumber \\}
\newcommand{\ba}[1]{\begin{array}{#1}}
\newcommand{\ea}{\end{array}}

\newcommand{\bs}[1]{\be \left\{ \ba{#1}}
\newcommand{\es}{\ea \right. \ee}

\title{
\begin{flushright}
 \mbox{\normalsize ITEP/TH-23/09}
\end{flushright}
\vskip 20pt
Complanart of system of polynomial equations}
\author{Vlasov Andrey D.}
\begin{document}
\hfill ITEP/TH-23/09

\bigskip

\centerline{\Large{
Complanart of system of polynomial equations
}}

\bigskip

\centerline{ Andrey Vlasov \footnote{Also at Moscow Institute for Physics and Technology and Institute of Astronomy of Russian Academy of Science \\ e-mail: vlasov.ad@gmail.com}}

\bigskip

\centerline{\it ITEP, Moscow, Russia}

\bigskip

\centerline{ABSTRACT}
In this paper we study polynomial maps of vector spaces $z_i \rightarrow A_i^{i_1i_2\cdots i_s}z_{i_1}z_{i_2}z_{i_3}\cdots z_{i_s}$  and their eigenvectors and eigenvalues. The new quantity called complanart is defined. Complanarts determine complanarity of solution vectors of systems of polynomial equations. Evaluation of complanart is reduced to evaluation of resultants. As in linear case, the pattern of eigenvectors defines the phase diagram of associated differential equation $\dot{z}_i=A_i^{i_1i_2\cdots i_s}z_{i_1}z_{i_2}z_{i_3}\cdots z_{i_s}$. Theory of such differential equations arise naturally as extension of Lyapunov's theory of stability for solutions of differential equations. The results of this work have a number of potential applications: from solving non-linear differential equations and calculating non-linear exponents to taking non-Gaussian integrals.  
\tableofcontents
\section{Introduction}
\subsection{Overview}
This paper is about non-linear algebra \cite{GKZ,INT}, which studies non-linear maps $z_i \rightarrow A_i^{i_1i_2\cdots i_s}z_{i_1}z_{i_2}z_{i_3}\cdots z_{i_s}$ and systems of polynomial equations. Non-linear algebra is a direct generalization of linear algebra \cite{G}. %The notion of resultant, and some other necessary objects/definitions are introduced in sect \ref{termnot}.
 Developing non-linear algebra can help us to take non-Gaussian integrals, and can convert many calculations in physics in exact ones. It can also be applied to theory of stability of differential equations. %, for details see sect \ref{nongaussintsect}.
%  In sect \ref{sectsovp} the ways of calculating different symmetric combinations of the roots of polynomial equations, other that resultant and discriminant, are discussed.
 Instead of determinants, the main functions/objects of quantitative non-linear algebra are discriminants and resultants \cite{GKZ,INT}. Resultants determine solvability  of system of polynomial equations, and discriminant determines the degeneracy of non-linear form. In this paper we consider another, new measure of degeneracy of system of polynomial equations, called complanart. Complanart determines, whether there are $n$ complanar roots of system, for details see sect \ref{sovpintrod}, and sect \ref{subsovpad}. In sect \ref{eigenvectors} the problem of finding non-linear eigenvectors/eigenvalues is discussed. In sect \ref{sectdiffur} the applications of all derived methods to the theory of polynomial differential equations are presented. The equations of such type arise in some degenerate cases in the theory of stability. In sect \ref{example} all derived techniques and methods are illustrated on one example of quadratic map of two variables.
\subsection{Symmetric combinations of the roots}
It is well known (see, for example, \cite{Galois}), that all polynomial symmetric combinations of the roots of system of $n-1$ homogeneous equations of $n$ variables in principle can be expressed polynomially in the coefficients of these equations. But explicit expressions for the roots themselves, with separate expression for each root, exist only in simple cases. "Explicit" means the expression, involving only arithmetic roots and algebraic operations.  For example, for one equation of two homogeneous variables such expressions exist only for degrees from 1 to 4.
% But, according to Galois theory \cite{Galois}, \textbf{polynomial symmetric combinations of roots of the system are always expressed polynomially on the coefficients of the system.} 
So, it is sometimes easier and more convenient to study the system without finding all roots, but by studying symmetrical combinations of the roots. For example, discriminant of a form determines whether the form is degenerate or not, resultant (see sect.\ref{termnot} and \cite{INT,GKZ}) of a system of polynomial equations determines whether the system has non-trivial solution. One more example of successful application of this approach to higher discriminants of polynomials described in \cite{CoiRoots}. Known difficulties of this way of analysis is that there has been developed no clear methods of obtaining expression for arbitrary symmetric combinations of the roots through coefficients of the system. In this paper new symmetric combination of the roots, called complanart, is considered. The evaluation of this quantity is reduced to evaluation of resultants. Calculation of elementary symmetric polynomials of roots, namely the generalization of Vieta formulas, is also reduced to evaluation of resultants, see \ref{Vieta}.
\subsubsection{Complanart \label{sovpintrod}}
Complanart is a symmetric combination of the roots of a system of polynomial equations. Let $f_1(x),\cdots,f_{n-1}(x)$ be $n-1$ homogeneous polynomials of arbitrary degrees $r_i$ of $n$ variables $x_1,\cdots,x_n$, and let $\Lambda^{(1)},\cdots,\Lambda^{(N)}$ are roots of the system:
\be
f_1(x)=0 \sls
\vdots 
% \label{systforvietintro} 
\nonumber  \\
f_{n-1}(x)=0 \nonumber
\ee
This system has in general case $N=r_1r_2\ldots r_{n-1}$ roots with at least one non-zero component up to overall rescaling, see, for example \cite{Galois}. Double roots are counted and repeated two times, roots of third order - three times etc. Complanart equals:
\be \boxed{
C=\prod_{\stackrel{i_1,\cdots,i_n=1}{i_1<i_2<\cdots<i_n}}^N\left(\eps^{j_1j_2\cdots j_n}\Lambda^{(i_1)}_{j_1}\Lambda^{(i_2)}_{j_2}\cdots\Lambda^{(i_n)}_{j_n}\right)^2
}
\ee
\textbf{Complanart equals $0$ iff there is a set of $n$ complanar roots}, i. e. there is a set of roots $\Lambda^{(i_1)},\Lambda^{(i_2)},\cdots,\Lambda^{(i_n)}$ without pair of equal indices $i_1\ne i_2\ne i_3\ne \cdots \ne i_n$ satisfying \\ $\eps^{j_1j_2\cdots j_n}\Lambda^{(i_1)}_{j_1}\Lambda^{(i_2)}_{j_2}\cdots\Lambda^{(i_n)}_{j_n}=0$. In particular, if there is at least one multiple root, complanart also equals zero. In the case $n=2$ complanarity of vectors means their collinearity, %. In this case there are only one equation and two variables
 so complanart reduces to ordinary discriminant of polynomial. If the number of distinct roots is less than $n$, complanart equals 1. For example, complanart equals 1 if all equations are linear equations. The method of evaluating complanart is described in sect.\ref{antisymmprod}. More details about complanarts can be found in sect \ref{subsovpad}.

\subsection{Eigenvectors and eigenvalues \label{introdeig}}
Eigenvector and eigenvalue of linear maps are known from linear algebra, and have direct analogues in non-linear algebra.
%The next features, known from linear algebra, are eigenvector and eigenvalue.
 Non-zero vector $z_i$ is called eigenvector of $A$, if it satisfies (with some $\lambda$):
\be
A_i^{i_1i_2\cdots i_s}z_{i_1}z_{i_2}z_{i_3}\cdots z_{i_s}=\lambda(z)z_i  \nonumber
%\label{eigenvector}
\ee
$\lambda$, a polynomial of degree $s-1$, is called eigenvalue. In non-linear case one can at first find eigenvectors and then find eigenvalues by solving linear equations, see \ref{unitaryeig}.
% It turns out, that in non-linear case the normalization of eigenvectors with nonzero eigenvalue is important, unlike linear case.
%As we will see in sect \ref{unitaryeig}, 
Since to find eigenvalues one have to solve linear on $\lambda$ equations, the set of eigenvalues is a union of planes in the space of all polynomials of degree $s-1$. This statement can be reformulated using characteristic polynomial of the map, namely:
\be
 Ch_A(\lambda)\equiv R\{A_i(z)-\lambda(z)z_i\} \nonumber
 % \label{charmnog}
\ee
$Ch_A(\lambda)=0$ iff $\lambda$ is an eigenvalue of $A$.
% The statement, that set of eigenvalues is a union of planes in the space of all polynomials of degree $s-1$, is equivalent to the statement that characteristic polynomial possesses decomposition on linear on the coefficients of $\lambda$ factors.
 Characteristic polynomial possesses decomposition on linear in the coefficients of $\lambda$ factors, since the set of eigenvalues is a union of planes in the space of all polynomials of degree $s-1$. This decomposability was firstly stated in \cite{INT}, but without a proof. The proof will be given in \ref{decomp}\\
% $Ch_A$ is non-homogeneous polynomial of coefficients of $\lambda$
Finding eigenvectors is reduced to solving the system of $n$ homogeneous equations of $n+1$ variables. Such systems possess complanart. Complanart can be used to determine whether the system has complanar eigenvectors. Here "complanar" means complanarity of vectors in extended space, i. e. in space with additional homogenizing variable, see \ref{sovpappl}. \\
The number of eigenvectors of non-degenerate map equals $c_{n|s}=\frac{s^n-1}{s-1}$, if there is no degenerations such as coinciding eigenvectors or the case when the map is unit map. The formula for $c_{n|s}$ was stated in \cite{INT}, but from considerations for diagonal maps. In sect \ref{numeig} this formula is derived in general case.
%Then one find solutions to $Ch_A(\lambda)=0$, substitute these $\lambda$'s in (\ref{eigenvector}), and find eigenvectors. In s.\ref{unitaryeig} we show that this way is not convenient, and it doesn't take advantage of an important property of characteristic polynomial - its factorizability (it is worth emphasizing, that not {\it any} polynomial of 2 and greater  number of variables can be factorized on {\it linear} on these variables factors; but characteristic polynomial as polynomial on the coefficients of $\lambda$ always possess such decomposition). Factorizability was discovered in \cite{INT}. The explanation of this fact, the nature and the expressions of the factors are discussed in s.\ref{decomp} There is a difference between linear and nonlinear cases, namely in the linear case eigenvalue was a polynomial of degree 0, i. e. a number. Nonlinear eigenvalues can be rescaled to $1$ or $0$, using known eigenvectors (see discussion s.\ref{unitaryeig}). 
%instead properly normalized eigenvectors content all the information about the map.
% We develop the method of finding eigenvectors/eigenvalues, and describe, how many eigenvectors will be obtained, and which degenerations can happen with them (see s.\ref{numeig}). Using complanart, one can understand (without finding eigenvectors/eigenvalues), whether a map has complanar eigenvectors or not (see s.\ref{sovpappl}). 
\subsection{Applications of our approach}
%Now let us consider the application of the developed techniques to some problems beyond this work.
\subsubsection{Nonlinear differential equations \label{nonlindiffur}}
The eigenvectors of a map $A_i^{i_1i_2\cdots i_s}$ entirely determine the phase diagram of the system of differential equations:
%Eigenvectors of a non-linear map are very important when considering such differential equations:
\be
\dot{z}_i=A_i^{i_1i_2\cdots i_s}z_{i_1}z_{i_2}z_{i_3}\cdots z_{i_s} \label{diffur}
\ee
If initial condition is proportional to an eigenvector, the solution is very simple, see sect \ref{eigdiffur}.
\paragraph{Application in the theory of stability}
The equations of type (\ref{diffur}) arise naturally when considering the stability of the stationary point of system of differential equations. Consider a system of ordinary differential equations:
\be
\left\{
\ba{c}
\dot{y_1}=f_1(y_1,y_2,\cdots,y_n) \\
\dot{y_2}=f_2(y_1,y_2,\cdots,y_n) \\
\vdots \sls
\dot{y_n}=f_n(y_1,y_2,\cdots,y_n)
\ea
\right.
\ee
Let $y^{(0)}_1,\cdots,y^{(0)}_n$ be a stationary point of this system: $f_1(y^{(0)}_1,\cdots,y^{(0)}_n)=0,\cdots,f_n(y^{(0)}_1,\cdots,y^{(0)}_n)=0$. Denote now $y'_i=y_i-y^{(0)}_i$. To analyse stability of this point and some quantitative properties of this stability/instability one can 
%$(y_i)_{new}=(y_i)_{old}-y^{(0)}_i$ 
expand $f_1,\cdots,f_n$ around the stationary point:
\be
\dot{y}'_i=\left.\left(\frac{\p f_i}{\p y_j}\right)\right|_{0,0,\cdots,0}y'_j+\frac{1}{2!}\frac{\p^2 f_i}{\p y_j \p y_k}y'_jy'_k+\cdots
\ee
The first term in expansion $\left.\left(\frac{\p f_i}{\p y_j}\right)\right|_{0,0,\cdots,0}y'_j$ is linear map, the second $\frac{1}{2!}\frac{\p^2 f_i}{\p y_j \p y_k}y'_jy'_k$ is homogeneous quadratic map, and so on. The case when there {\it is} a non-degenerate linear term in this expansion is well-known, this is a subject of consideration of Lyapunov theory of stability (see \cite{Lapunov}) with its Lyapunov's indices equal to eigenvalues of the matrix $\left.\left(\frac{\p f_i}{\p y_j}\right)\right|_{0,0,\cdots,0}$. But for some differential equations linear term in this expansion vanishes, $\left.\left(\frac{\p f_i}{\p y_j}\right)\right|_{0,0,\cdots,0}=0$. In such cases it is necessary to consider the terms in expansion of higher degrees. If only the term of the lowest degree is considered, one arrives to the system of type (\ref{diffur}). The physically motivated example of system of differential equations with vanishing linear term will be given in sect.\ref{examplevan}. The discussion of stability/instability of points with vanishing linear term is given in sect.\ref{uslstab}. For example, if the resultant of the main non-linear term does not equal to zero, the point is unstable in its complex vicinity. If main non-linear term has a real eigenvector, this point is unstable in its real vicinity.
%  Some ideas about stability/instability of equations of type (\ref{diffur}) are written in sect.\ref{eigdiffur}.
% One of results of our work is, that, in the case when linear term vanishes, any stationary point is unstable in its complex vicinity (see s.\ref{eigdiffur}).
% In the solution of the equations (\ref{diffur}), the eigenvectors play very important role and entirely determine the phase diagram of the system. Also if the initial condition is proportional to an eigenvector, the solution is very simple and intuitive.
\subsubsection{Non-Gaussian integrals \label{nongaussintsect}}
The formula for Gaussian integral and obtained from it Wick theorem are widely used in modern science. In many cases when it is necessary to calculate something Gaussian integrals are used. Gaussian integral is the integral:
\be
Z(J)\equiv\int\limits_{-\infty}^{+\infty} e^{-A^{ij}x_ix_j+J^ix_i} dx_1\cdots dx_n=\sqrt{\frac{\pi^n}{det A}}e^{-\frac{\left(A^{-1}\right)_{ij}J^iJ^j}{4}} % \nonumber
\label{GaussInt}
\ee
For example, Feynman diagram technique uses them. In it this formula is extended from ordinary to functional integrating. One is really interested in calculating such quantities:
\be
<\phi(x_1)\phi(x_2)\cdots\phi(x_k)>=\frac{\left.\int D\phi( \phi(x_1)\phi(x_2)\cdots\phi(x_k))e^{\frac i\hbar\int(L(\phi)+J\phi)d^4x}\right|_{J=0}}{\left.\int D\phi e^{\frac i\hbar\int(L(\phi)+J\phi)d^4x}\right|_{J=0}} %\nonumber\\ 
 \label{nongauskor} % \\
%Z(J)\equiv \int D\phi( \phi(x_1)\phi(x_2)\cdots\phi(x_k))e^{\frac i\hbar\int(L(\phi)+J\phi)d^4x} \nonumber
% \label{nongaus}
\ee
which are called corellators. $\phi$ is a field (or fields), $L(\phi)$ is a Lagrangian of this field(s) and $J(x)\phi(x)$ is called source of the field.
% (и как же по-умному называетÑ?Ñ? $J$?).
% How does one calculate these integrals with Feynman technique? First, let us consider how to calculate
If $L(\phi)$ contains only terms, quadratic on $\phi$, for example $L(\phi)=\p_i\phi\p^i\phi-m^2\phi^2$ - a lagrangian for free scalar massive field, these quantities are calculated as follows:
%Such integrals with quadratic Lagrangian (i. e. $L(\phi)$ contains only terms, quadratic on $\phi$, for example $L(\phi)=\p_i\phi\p^i\phi-m^2\phi^2$ - a lagrangian for free scalar massive field) are calculated as follows:
\be
Z(J)\equiv\int D\phi( \phi(x_1)\phi(x_2)\cdots\phi(x_k))e^{\frac i\hbar\int(L_0(\phi)+J\phi)d^4x}= \frac{const}{\sqrt{det(L_0)}}e^{-L_0^{-1}(J,J)} \label{gausobob} \\
<\phi(x_1)\phi(x_2)\cdots\phi(x_k)>_{free}=\frac{\int D\phi( \phi(x_1)\phi(x_2)\cdots\phi(x_k))e^{\frac i\hbar\int(L_0(\phi))d^4x}}{\int D\phi e^{\frac i\hbar\int(L_0(\phi))d^4x}}= \sls
=\frac{\left.\left(\frac{\p}{\p J(x_1)} \frac{\p}{\p J(x_2)}\cdots \frac{\p}{\p J(x_m)} Z(J)\right)\right|_{J=0} }{Z|_{J=0}} \label{gausobobres}
\ee
The formula (\ref{gausobob}) is the direct generalization of (\ref{GaussInt}) to the case of functional integrating, and determinant in it is so-called functional determinant. $L_0^{-1}$ is called propagator of the field $\phi$. This result in another form is also called the Wick theorem. Now, to calculate (\ref{nongauskor}) one just expands non-quadratic terms in the exponent and calculates only correlators in the free theory (i. e. in the theory with quadratic Lagrangian).
% All the considerations are right if there are several fields.
% This consideration was done here only to show that Feynman technique is nothing but one of the method of perturbative calculation of non-Gaussian integrals (that is (\ref{nongauskor})).
The quantities (\ref{nongauskor}) in this approach are calculated perturbatively. It is more preferable to calculate them non-perturbatively, {\it exactly}. To solve this problem, it is necessary to evaluate integrals:
\be
 \int e^{(J^ix_i+A^{ij}x_ix_j+B^{ijk}x_ix_jx_k+\cdots)}dx_1dx_2\cdots dx_n \label{nongaussint}
\ee
{\it in the limit $n\rightarrow \infty$} ($n$ is the number of $x_i$). Dots in the exponent substitute parts of the Lagrangian of higher degrees. The integrals of type (\ref{nongaussint}) are called non-Gaussian integrals.
%: if a lagrangian contains only quadratic and cubic terms, the written part is all the formula, if the terms of higher degrees are also present, we will need to complete exponent.
% Now the scientists are still far from the solution of this problem: we can evaluate only some particular cases of non-gaussian integrals. 
Non-Gaussian integrals are also studied in \cite{QFT}. Now there are no simple methods of evaluating such integrals, and the expressions for these integrals were obtained in \cite{Integrals} only for several simple cases. The expressions for these integrals depend on the number of variables (unlike the gaussian integral), therefore it is not evident how they behave in the limit $n\rightarrow\infty$.
% For further reading in current progress see \cite{Integrals}.
In \cite{Integrals} it was shown, that discriminants of non-linear forms play an important role in the evaluation of the non-Gaussian integrals, e. g. they control singularities of these integrals. 
%The possible approach, alternative to one in \cite{Integrals}, is to use 
%One can try to calculate Gaussian integral using 
The possible approach to calculating non-Gaussian integrals is to use some form of canonical representation of non-linear form. In general case, non-linear form cannot be brought to diagonal representation, but the free parameters of transformations can be used to fix up some coefficients of the form. A possible variant of such representation of the maps ($A^{ji}_k,A_i^{jkm},\dots$) under $GL_n$ action is presented in sect \ref{canrepres}. Under the action of $SO_n$ the canonical representations of forms  $A_{ijk},A_{jkim},\dots$ and for maps $A^{ji}_k,A_i^{jkm},\dots$ are the same, therefore studying canonical representation of map under $SO_n$ can help us to evaluate non-Gaussian integrals.
% However, to calculate non-Gaussian integrals, it is needed the canonical representation of non-linear forms, e. g. $A_{ijk},A_{jkim},\dots$. The classification of maps and of forms under $SL_n$ is the same.
% How this paper could help us to evaluate non-gaussian integral? We can evaluate gaussian integral because we know how to bring a quadratic form to canonical view by orthogonal transformations. But, if we consider only orthogonal transformations, quadratic form and linear map are the same thing (quadratic form has form of matrix $A^{ij}$, and linear map has form of matrix $A^i_j$, and we can raise/lower index only under $SO_n$ symmetry, but not $GL_n$). So, if some theory of canonical form of non-linear maps under the action of $SO_n$ will be developed, this theory will help us to evaluate non-gaussian integrals. For example, quadratic map is dual (in the above sense) to cubic form ($A_i^{jk}\leftrightarrow A^{ijk}$), the cubic map is dual to the form of degree 4 ($A_i^{jkm}\leftrightarrow A^{ijkm}$) and so on. In this paper we consider only $GL_n$ canonical form. This is because of nonlinear maps have so many parameters, and $SO_n$ group has so few parameters to make the representation of nonlinear map convenient and useful. Even $GL_n$ canonical representation of the map is not such easy thing, and one example of its unvestigation is given in s.\ref{example}
\subsection{Terms and notations \label{termnot}}
\subsubsection{Homogeneous and non-homogeneous equations \label{nonhom}}
Polynomial $f(x_1,\dots,x_n)$ in variables $x_1,\dots,x_n$ is called {\it homogeneous polynomial}, if for any $\lambda\ne 0:\quad f(\lambda x_1,\dots,\lambda x_n)=\lambda^df(x_1,\dots,x_n)$. Non-negative integer $d$ is called the degree of $f$. Any homogeneous polynomial can be made non-homogeneous by dividing it by one of the variables to power d, for example, by $(x_n)^d$. After such division, the ratios $x_1/x_n,x_2/x_n,\dots,x_{n-1}/x_n$ can be taken as new variables: $y_1=x_1/x_n,y_2=x_2/x_n,\dots,y_{n-1}=x_{n-1}/x_n$, so the number of variables was decreased by one. The variables $y_1,\dots,y_{n-1}$ are called non-homogeneous variables, and $x_1,\dots,x_n$ are called homogeneous variables. In this paper we will denote by $\Lambda^{(i)}_j$ j-th component of i-th root of a system of equations in homogeneous variables, and by $\lambda^{(i)}$ i-th root of one equation in one non-homogeneous variable. For example, if the equation were $a_0 (x_2)^d+a_1(x_2)^{d-1}x_1+\cdots+a_{d-1}x_2(x_1)^{d-1}+a_d(x_1)^d=0$, then $\Lambda^{(i)}_1$ is $x_1$-component of i-th solution, $\Lambda^{(i)}_2$ is $x_2$-component of i-th solution. This equation in non-homogeneous variable $z=\frac{x_1}{x_2}$ states $a_0+a_1z+\cdots+a_{d-1}z^{d-1}+a_dz^d=0$. $\lambda^{(i)}$ is i-th root of this equation. In sect \ref{sectsovp} we consider only homogeneous functions of $\Lambda^{(i)}_j$, because all $\Lambda^{(i)}_j$ can be simultaneously rescaled and remain the solution of the system. Symmetrical combination of roots $\Lambda^{(i)}$ is an expression, which does not change under swapping of any pair $\Lambda^{(i)}$ and $\Lambda^{(j)}$. 
% Under symmetrical combination of roots $\Lambda^{(i)}$ is meant the expression, which does not change under swapping of any pair $\Lambda^{(i)}$ and $\Lambda^{(j)}$. 
%\footnote{ For example, if the equation were $a_0 x^d+a_1x^{d-1}y+\cdots+a_{d-1}xy^{d-1}+a_dy^d=0$, then $\Lambda^{(i)}_1$ is $x$-component of i-th solution, $\Lambda^{(i)}_2$ is $y$-component of i-th solution. This equation is called homogeneous equation (sum of degrees of $x$ and $y$ in some term is equal to that of another term). But we can solve this equation in other form: divide two sides of equality by $x^d$, then denote $z=\frac{y}{x}$, and we obtain the equation: $a_0+a_1z+\cdots+a_{d-1}z^{d-1}+a_dz^d=0$. $z$ is called nonhomogeneous variable, and $\lambda^{(i)}$ is i-th root of non-homogeneous equation}
\subsubsection{Maps and resultants \label{resultant}}
We study homogeneous polynomial maps of vector spaces:
\be
z_i \rightarrow A_i^{i_1i_2\cdots i_s}z_{i_1}z_{i_2}z_{i_3}\cdots z_{i_s} \nonumber
\ee
 The degree of the map is denoted by $s$. If $s=1$, we get ordinary linear maps, which are the objects of consideration of standard course of linear algebra \cite{G}. Many objects of linear algebra can be generalized to describe the non-linear case.
%, and we will define now important part of them, which we will use in our work. 
First, consider the system of equations 
\be
A_i^{i_1i_2\cdots i_s}z_{i_1}z_{i_2}z_{i_3}\cdots z_{i_s}=0 \nonumber % \label{systforres}
\ee
This is a system of $n$ homogeneous equations of $n$ homogeneous variables or $n-1$ non-homogeneous variables.
% We can divide each equation by, for example, $z_1$, and then this will be a system of $n$ equations of $n-1$ variables $z_2/z_1,\dots,z_n/z_1$, called non-homogeneous variables.
  Such system in general case has no non-trivial solutions at all. Non-trivial solution is a solution with at least one component being non-zero. For such solution to exist, the coefficients of the system must satisfy one relation, because the number of variables minus the number of equations equals one.
% System of equations (\ref{systforres}) is the system of .
% It is known, that non-zero solution exists if one relation ($n-(n-1)=1$) is satisfied by the the coefficients of this system.
This relation is
\be
 R\{A\}=0 \nonumber
\ee
where $R\{A\}$ is polynomial of the coefficients of $A$, called resultant or hyperdeterminant of $A$. If $s=1$, i. e. $A$ is the linear map, the resultant reduces to ordinary determinant of matrix. Resultants play an increasing role in modern mathematics and physics. See, for example \cite{INT,GKZ,Cay,D1,D2} for overview, \cite{recent,robots} for applications in physics and engineering, \cite{UFN2} for application in string theory and \cite{Complex,ContInt} for computational methods. By non-degenerate map we mean map with non-vanishing resultant.
\section{Complanart and symmetric combinations of the roots \label{sectsovp}}
\subsection{ Resultant and generalization of Vieta formulas \label{Vieta}}
%How Vieta formulas and resultants are related with each other? What do we mean by Vieta formulas?
% which express symmetrization on vector indices of the product on all projectively different 
Let $f_1(x),\cdots,f_{n-1}(x)$ be $n-1$ homogeneous polynomials of arbitrary degrees $r_i$ of $n$ variables $x_1,\cdots,x_n$. It is well known, that the system of equations:
\be
\left\{
\ba{c}
f_1(x)=0  \\ %\sls
\vdots  \label{systforviet} \\
f_{n-1}(x)=0 % \nonumber
\ea
\right.
\ee
 in general case has $N=r_1r_2\cdots r_n$ projectively-inequivalent solutions, see \cite{Galois}. Sometimes they may coincide, so $N$ should account for multiplicity. For example, {\it one} polynomial of 2 homogeneous variables or 1 non-homogeneous variable has exactly $r_1$ projectively-inequivalent solutions (with multiplicities). Denote different solutions of (\ref{systforviet}) by $\Lambda^{(1)},\Lambda^{(2)},\cdots,\Lambda^{(N)}$. Each solution is also a vector, so $\Lambda^{(i)}_j$ is a j-th vector component of i-th solution vector. Thus, Vieta formulas:
\be
% symm_{j_1,j_2,\cdots,j_N}\{\Lambda^{(1)}_{j_1}\Lambda^{(2)}_{j_2}\cdots\Lambda^{(N)}_{j_N}\}=V_{j_1\cdots j_N}
{\rm symm}_{\mu_1,\cdots,\mu_N}\{\Lambda^{(\mu_1)}_{j_1}\Lambda^{(\mu_2)}_{j_2}\cdots\Lambda^{(\mu_N)}_{j_N}\}\equiv\sum_{\sigma \in P_N}\Lambda^{(\sigma(1))}_{j_1}\Lambda^{(\sigma(2))}_{j_2}\cdots\Lambda^{(\sigma(N))}_{j_N}=V_{j_1\cdots j_N}  %\nonumber
 \label{Vietafor}
\ee
%where $V_{j_1\cdots j_N}$ is some polynom of coefficients of $f_1,\cdots,f_{n-1}$.
 $V_{j_1\cdots j_N}$ is homogeneous polynomial of coefficients of $f_i$. The degree of $V_{j_1\cdots j_N}$ in the coefficients of i-th polynomial $f_i$ equals $N/r_i=r_1\cdots r_{i-1}r_{i+1}\cdots r_i$, see \cite{INT}. In some particular cases the formula for $V_{j_1\cdots j_N}$ may be obtained from simple considerations (see for detailed discussion and examples \cite{INT}), but we consider now a general method of calculating it. This method uses a Poisson product formula for resultant, see e. g. \cite{GKZ,ContInt}. Add one more homogeneous polynomial $g(x)$ of degree $r$ of the same variables $x_1,\cdots,x_n$. Then the system of equations
\be 
\left\{
\ba{c}
f_1(x)=0 \label{systforres} \\ %\sls
\vdots \nonumber \\ % \label{systforviet} \\
f_{n-1}(x)=0 \\ % \nonumber
g(x)=0 \nonumber
\ea
\right.
\ee
will possess a resultant. Poisson product formula states:
\be
 R\{f_1,\cdots,f_{n-1},g\}=C\prod_{i=1}^N g(\Lambda^{(i)}) \label{Poisson}
\ee
$C$ is a constant, depending on the normalization of the roots; we can normalize the roots in such a way, that $C$ will be equal 1. This formula has the following meaning: the system has non-zero solution iff $g$ equals zero on one of the roots of other functions. With the help of this formula, one easily obtains the tensor $V_{j_1\cdots j_N}$. One can substitute for $g$ linear function: $g(x)=g^ix_i$, then calculate the resultant of the system (\ref{systforres})
%$f_1(x)=0,\cdots,f_{n-1}(x)=0,g(x)=0$
, and (as it is easily seen from Poisson product formula):
\be
 V_{j_1\cdots j_N}= {\rm the \quad coefficient\quad before\quad} g^{j_1}\cdots g^{j_N}\quad {\rm in}\quad R\{f_1,\cdots,f_{n-1},g\} \nonumber
\ee
\subsection{Symmetric combinations \label{antisymmprod} }
The formula (\ref{Poisson}) is easily generalized. For example, one needs to calculate:
\be
 P_g=\prod_{i,j=1}^N g(\Lambda^{(i)},\Lambda^{(j)}) \label{Pg}
\ee
At first one calculates the resultant of the system:
\be
\left\{
\ba{c}
g(x,y)=0 \\
f_1(x)=0 \\
\vdots \\
f_{n-1}(x)=0 \nonumber
\ea
\right.
\ee
on the variables $x_1,\cdots,x_n$, treating $y_1,\cdots,y_n$ as parameters. This resultant we will denote \\ $R_x\{g,f_1,\cdots,f_{n-1}\}$. It is still a polynomial in the variables $y_1,\cdots,y_n$. Then one computes the resultant of the system:
\be
\left\{
\ba{c}
R_x\{g,f_1,\cdots,f_{n-1}\}(y)=0 \sls
f_1(y)=0 \sls
\vdots \sls
f_{n-1}(y)=0
\ea
\right.
\ee
in variables $y_1,\cdots,y_n$. Now it is obvious, how to get
\be
\prod_{i,j,k=1}^N h(\Lambda^{(i)},\Lambda^{(j)},\Lambda^{(k)}) \nonumber \\
\prod_{i,j,k,m=1}^N u(\Lambda^{(i)},\Lambda^{(j)},\Lambda^{(k)},\Lambda^{(m)}) \nonumber \\
\vdots \nonumber
\ee
One should calculate resultant as many times, as there are different arguments in desired function of roots. 
But in sect \ref{subsovpad} expressions of such a type are needed:
%One could now want to calculate instead of (\ref{Pg}) such an expression:
%Now imagine, that we need in such an expression:
\be
P'_g\equiv\prod_{\stackrel{i,j=1}{i\ne j}}^N g(\Lambda^{(i)},\Lambda^{(j)}) \label{Pprimeg}
%=\left(\prod_{\stackrel{i,j=1}{i<j}}^N g(\Lambda^{(i)},\Lambda^{(j)})\right)^2
\ee
The difference between (\ref{Pg}) and (\ref{Pprimeg}) is, that there is a product over all {\it pairs of non-coincident indices } in (\ref{Pprimeg}), but in (\ref{Pg}) there is a product over {\it all} pairs of indices. It seems, that (\ref{Pprimeg}) can be calculated in this way:
\be
P'_g=\prod_{\stackrel{i,j=1}{i\ne j}}^N g(\Lambda^{(i)},\Lambda^{(j)})=\frac{\prod\limits_{i,j=1}^N g(\Lambda^{(i)},\Lambda^{(j)})}{\prod\limits_{i=1}^N g(\Lambda^{(i)},\Lambda^{(i)})} \label{otnosh}
\ee
%Both numerator and denominator in this expression we can evaluate. 
Numerator in this expression can be evaluated using the formula (\ref{Pg}), and the denominator can be evaluated using (\ref{Poisson}). But a problem can arise. If $g(x,y)$ is an {\it antisymmetric} function $g(x,y)=-g(y,x)$,
% For such a function
 both numerator and denominator of (\ref{otnosh}) are equal to zero. So $P'_g$ can not be evaluated straightforwardly using (\ref{otnosh}). We have found a way for evaluating $P'_g$ in this case. Let $g_0(x,y)$ - some antisymmetric linear polynomial on $x_1,\cdots,x_n,y_1,\cdots,y_n$. Let us consider $g(x,y)=g_0(x,y)+tg_1(x,y)$, where $g_1(x,y)$ is
\be
g_1(x,y)=\sum_{i,j=1}^nx_iy_j=\left(\sum_{i=1}^nx_i\right)\left(\sum_{i=1}^ny_i\right) \label{gone}
\ee
%It is obvious, that this choose of $g_1$ is not invariant under $GL_n$ action. It is still an open question - how to formulate this technique by invariant way. 
This expression seems strange because it is not invariant under $GL_n$ action. We will discuss it a bit later. So:
\be
P'_{g_0}\equiv\prod_{\stackrel{i,j=1}{i\ne j}}^N g_0(\Lambda^{(i)},\Lambda^{(j)})=\lim_{t\rightarrow 0}{\frac{\prod_{i,j=1}^N g(\Lambda^{(i)},\Lambda^{(j)})}{\prod_{i=1}^N g(\Lambda^{(i)},\Lambda^{(i)})}}= \sls
= \lim_{t\rightarrow 0}{\frac{\prod_{i,j=1}^N (g_0(\Lambda^{(i)},\Lambda^{(j)})+tg_1(\Lambda^{(i)},\Lambda^{(j)}))}{\prod_{i=1}^N (g_0(\Lambda^{(i)},\Lambda^{(i)})+tg_1(\Lambda^{(i)},\Lambda^{(i)}))}} \label{sovplimit}
\ee
For non-linear polynomials $g_0(x,y)$ there is the same technique, the tensor $g_1(x,y)$ should be chosen in other way. For example, for quadratic $g_0(x,y):$
\be
 g_1(x,y)=\sum_{i,j,k,m=1}^nx_ix_jy_ky_m=\left(\sum_{i=1}^nx_i\right)\left(\sum_{i=1}^nx_i\right)\left(\sum_{i=1}^ny_i\right)\left(\sum_{i=1}^ny_i\right) \label{gtwo}
\ee
This method works not only for antisymmetric function of two variables, but for any number of variables. This is the formula for three variables:
\be
P'_{g_0}\equiv\prod_{\stackrel{i,j,k=1}{i\ne j;i\ne k;k\ne j}}^N g_0(\Lambda^{(i)},\Lambda^{(j)},\Lambda^{(k)})=\lim_{t\rightarrow 0}{\frac{\left(\prod\limits_{i,j,k=1}^N g(\Lambda^{(i)},\Lambda^{(j)},\Lambda^{(k)})\right)\left(\prod\limits_{i=1}^N g(\Lambda^{(i)},\Lambda^{(i)},\Lambda^{(i)})\right)^2}{\left(\prod\limits_{i,j=1}^N g(\Lambda^{(i)},\Lambda^{(i)},\Lambda^{(j)})\right)^3}} \label{sovplimitthree}
\ee
Now $g_0(x,y,z)$ - homogeneous function, $g(x,y,z)=g_0(x,y,z)+tg_1(x,y,z)$. For linear $g_0$:
\be
 g_1(x,y,z)=\sum_{i,j,k=1}^nx_iy_jz_k \label{gthreevar}
\ee
These formulas work for $g_0$ of any degree and of any symmetry.
% If $g_0$, for example, is not antisymmetric function, it is sufficient to take numerator and denominator of the formula (\ref{sovplimit}) or (\ref{sovplimitthree}) at $t=0$ and do not take the limit, and 
If $g(x,y,z)$ is of other degree, $g_1$ should be chosen of the same degree. For example, for quadratic $g_0$ we would write:
\be
g_1(x,y,z)=\sum_{i,j,k,m,l,p}^nx_ix_jy_ky_mz_lz_p \label{gthree}
\ee
%The motivations of choosing as (\ref{gone}),(\ref{gtwo}),(\ref{gthree}) are:
If the limits (\ref{sovplimit}),(\ref{sovplimitthree}) can be evaluated, they do not depend on the choice of $g_1$. But sometimes, if we choose degenerate $g_1$ or simply $g_1$ with small number of non-zero components, both numerator and denominator of (\ref{sovplimit}) will be zero even at $t\ne 0$, and we will not be able to calculate the limit. The example of this phenomena  will be given in s.\ref{sovpexample}. Expressions (\ref{gone},\ref{gtwo},\ref{gthree}) are simply examples of non-degenerate maps with all components being nonzero. The example of application of the formulae (\ref{sovplimit}),(\ref{sovplimitthree}) is calculation of complanarts.
% in the case of two and three variables.
\subsection{Complanart \label{subsovpad}}
When the system (\ref{systforviet}) has coincident roots? In the case of two variables (when there is only one polynomial) the answer is given by discriminant of the polynomial. Discriminant equals:
\be
 D=\prod_{i,j=1,i<j}^{r_1}\left(\eps^{km}\Lambda^{(i)}_k\Lambda^{(j)}_m\right)^2=\prod_{j,i=1,i<j}^{r_1}(\lambda^{(i)}-\lambda^{(j)})^2
\ee
%Here $\Lambda^{(i)}_k$ is the same, that was such way denoted in previous subsections, $\lambda^{(i)}$ is i-th root of non-homogeneous equation
%\footnote{ For example, if the equation were $a_0 x^d+a_1x^{d-1}y+\cdots+a_{d-1}xy^{d-1}+a_dy^d=0$, then $\Lambda^{(i)}_1$ is $x$-component of i-th solution, $\Lambda^{(i)}_2$ is $y$-component of the soluiton. This equation is called homogeneous equation (sum of degrees of $x$ and $y$ in some term is equal to that of another term). But we can solve this equation in other form: divide two sides of equality by $x^d$, then denote $z=\frac{y}{x}$, and we obtain the equation: $a_0+a_1z+\cdots+a_{d-1}z^{d-1}+a_dz^d=0$. $z$ is called nonhomogeneous variable, and $\lambda^{(i)}$ is i-th root of non-homogeneous equation}
$r_1$ is the degree of a polynomial. The discriminant equals zero iff there is a pair of roots, in which one is proportional to another ($\Lambda^{(i)}\propto\Lambda^{(j)}, i\ne j$, homogeneous formulation), or two equal non-homogeneous roots ($\lambda^{(i)}=\lambda^{(j)}$, non-homogeneous formulation). %From now on, we will use only homogeneous formulation. 
In three-dimensional space, however, the proportionality of two vectors is defined by two conditions (e. g. $x_2/x_1=y_2/y_1$ and $x_3/x_1=y_3/y_1$). But there is a natural one condition for complanarity of {\it three} vectors: $\eps^{ijk}x_iy_jz_k=0$. If $\eps^{ijk}x_iy_jz_k=0$, three vectors $x,y,z$ lie in one plane, or they are simply complanar. So, we can formulate the condition of a system of equations to have three complanar roots:
\be
C\equiv\prod_{\stackrel{i,j,k=1}{i<j<k}}^N\left(\eps^{mlp}\Lambda^{(i)}_m\Lambda^{(j)}_l\Lambda^{(k)}_p\right)^2, \label{sovpthreevar} \\
C=0  \label{cond}
\ee
(squared for symmetry). The condition itself is (\ref{cond}). (\ref{sovpthreevar}) is some symmetrical polynomial of the roots of the system. To make it symmetric combination $\eps^{mlp}\Lambda^{(i)}_m\Lambda^{(j)}_l\Lambda^{(k)}_p$ was squared. This symmetric polynomial of roots we call complanart.
% (\ref{sovpthreevar}) is a symmetrical combination of the roots of the system, and we call it complanart. The square there makes this expression symmetric on different 
For two variables complanart reduces to ordinary discriminant, for four variables:
\be
C=\prod_{\stackrel{i,j,k,q=1,}{i<j<k<q}}^N\left(\eps^{mlpr}\Lambda^{(i)}_m\Lambda^{(j)}_l\Lambda^{(k)}_p \Lambda^{(q)}_r\right)^2
\ee
and for $n$ variables:
\be
C=\prod_{\stackrel{i_1,\cdots,i_n=1}{i_1<i_2<\cdots<i_n}}^N\left(\eps^{j_1j_2\cdots j_n}\Lambda^{(i_1)}_{j_1}\Lambda^{(i_2)}_{j_2}\cdots\Lambda^{(i_n)}_{j_n}\right)^2
\ee
The complanart has degree $2nC_N^n=2n\frac{N!}{(N-n)!n!}$ on $\Lambda$, since every factor in product has degree $2n$ and there are $C_N^n$ factors. Each $\Lambda\propto\prod\limits_{i=1}^{n-1}(a_i)^{1/r_i}$, where $a_i$ denotes coefficients of $f_i$ (see (\ref{Vietafor})). So:
\be
C\propto \prod_{i=1}^{n-1}(a_i)^{2C_{N-1}^{n-1}\frac{N}{r_i}} 
\ee
i. e. $deg_{a_i}C=2C_{N-1}^{n-1}\frac{N}{r_i}$. If $n=2$, then complanart is discriminant, the number of solutions $N=r$ , and $\quad deg_{a}C=2\frac{(r-1)!}{1!(r-2)!}=2(r-1)$. It is well-known expression for degree of discriminant. 
\subsubsection{Evaluation of complanarts}
Complanart is just a product of values of antisymmetric function over sets of all different $n$ roots.
% sets of $n$ roots without any equal numbers of roots in the set.
  The method of evaluation of such quantities is given in \ref{antisymmprod}. It is necessary just to take for $g_0$ absolutely antisymmetric $\eps$-tensor of appropriate dimension. %Here are the formulae one more time:
The formula for two variables:
\be
C=\lim_{t\rightarrow 0}{\frac{\prod_{i,j=1}^N g(\Lambda^{(i)},\Lambda^{(j)})}{\prod_{i=1}^N g(\Lambda^{(i)},\Lambda^{(i)})}}=\lim_{t\rightarrow 0}{\frac{\prod_{i,j=1}^N (\eps(\Lambda^{(i)},\Lambda^{(j)})+tg_1(\Lambda^{(i)},\Lambda^{(j)}))}{\prod_{i=1}^N (\eps(\Lambda^{(i)},\Lambda^{(i)})+tg_1(\Lambda^{(i)},\Lambda^{(i)}))}}\label{sovptwo} 
\ee
For three variables:
\be
C^3=\lim_{t\rightarrow 0}{\frac{\left(\prod_{i,j,k=1}^N g(\Lambda^{(i)},\Lambda^{(j)},\Lambda^{(k)})\right)\left(\prod_{i=1}^N g(\Lambda^{(i)},\Lambda^{(i)},\Lambda^{(i)})\right)^2}{\left(\prod_{i,j=1}^N g(\Lambda^{(i)},\Lambda^{(i)},\Lambda^{(j)})\right)^3}} \label{sovpthree} \\
g(x,y,z)=\eps(x,y,z)+tg_1(x,y,z) \nonumber
\ee
For four variables:
\be
C^{12}=\lim_{t\rightarrow 0}\frac{\prod\limits_{i,j,k,m=1}^Ng(\Lambda^{(i)},\Lambda^{(j)},\Lambda^{(k)},\Lambda^{(m)})\left(\prod\limits_{i,j=1}^Ng(\Lambda^{(i)},\Lambda^{(i)},\Lambda^{(i)},\Lambda^{(j)}) \right)^8  \left(\prod\limits_{i,j=1}^Ng(\Lambda^{(i)},\Lambda^{(i)},\Lambda^{(j)},\Lambda^{(j)})\right)^3  }{\left(\prod\limits_{i,j,k=1}^Ng(\Lambda^{(i)},\Lambda^{(i)},\Lambda^{(j)},\Lambda^{(k)}) \right)^6 \left(\prod\limits_{i=1}^Ng(\Lambda^{(i)},\Lambda^{(i)},\Lambda^{(i)},\Lambda^{(i)})\right)^6 } \label{sovpfour} \\
g(x,y,z,u)=\eps(x,y,z,u)+tg_1(x,y,z,u) \nonumber
\ee
%In formula for three variables (\ref{sovpthree}) $g(x,y,z)=\eps(x,y,z)+tg_1(x,y,z)$, in formula for four variables (\ref{sovpfour}) $g(x,y,z,u)=\eps(x,y,z,u)+tg_1(x,y,z,u)$.
%The formula for arbitrary number of variables will be derived (а может бытÑ?, и нет) in appendix ... 
There is one more fact to explain. Why for three variables the limit yields to $C^3$, and for four variables it yields for $C^{12}$? It is because the formulae (\ref{sovptwo}),(\ref{sovpthree}),(\ref{sovpfour}) give us the following expressions:
\be
\prod_{\stackrel{i,j=1}{i\ne j}}^N \eps(\Lambda^{(i)},\Lambda^{(j)})\quad {\rm instead\quad of\quad}  \prod_{\stackrel{i,j=1}{i<j}}^N (\eps(\Lambda^{(i)},\Lambda^{(j)}))^2 \\
\prod_{\stackrel{i,j,k=1}{i\ne j;i\ne k;j\ne k}} \eps(\Lambda^{(i)},\Lambda^{(j)},\Lambda^{(k)})\quad {\rm instead\quad of\quad } \prod_{\stackrel{i,j,k=1}{i<j<k}}^N (\eps(\Lambda^{(i)},\Lambda^{(j)},\Lambda^{(k)}))^2 \\
\vdots \nonumber
\ee
In the case $n=2$ these two things coincide (because there is 2 permutations and accounting for them adds necessary squaring). In case $n=3$ there are $n!=6$ permutations, but a square is again needed - so appears $C^3$. In case of arbitrary $n$ this procedure yields $C^{\frac{n!}{2}}$. \\
It is now a simple exercise to write the formula for complanart analogous to (\ref{sovptwo}),(\ref{sovpthree}),(\ref{sovpfour}) in any particular dimension: it is necessary only that all the factors, appeared in $\prod\limits_{i_1,i_2,\cdots,i_n=1}^Ng(\Lambda^{(i_1)},\Lambda^{(i_2)},\cdots,\Lambda^{(i_n)})$ with some two or more $i$ coincident will be cancelled by factors with explicitly equal $i$, for example $\prod\limits_{i_1,i_2,\cdots,i_n=1}^Ng(\Lambda^{(i_1)},\Lambda^{(i_1)},\Lambda^{(i_3)},\cdots,\Lambda^{(i_n)})$ etc.
\subsubsection{When complanart is equal to 1?}
Complanart measures linear dependence of $n$ distinct roots of polynomial system of equations (\ref{systforviet}). Such system of equations in general case have $N=r_1r_2\cdots r_{n-1}$ projectively-nonequivalent roots, where $r_1,r_2,\cdots,r_{n-1}$ are degrees of equations. But what happens when $N<n$? It means, that there are {\it no} $n$ distinct roots, so no $n$ distinct roots can be complanar. Therefore the complanart of such a system is equal to non-zero constant, i. e. it {\it does not} depend on the coefficients of the system. What is this constant? Let us consider the simplest example: one linear equation, for example $ax_1+bx_2=0$. It has one solution: $\Lambda^{(1)}_1=-b,\Lambda^{(1)}_2=a$. Complanart equals:
\be
C=\lim_{t\rightarrow 0}{\frac{\prod_{i,j=1}^N g(\Lambda^{(i)},\Lambda^{(j)})}{\prod_{i=1}^N g(\Lambda^{(i)},\Lambda^{(i)})}}=\lim_{t\rightarrow 0}{\frac{\prod_{i,j=1}^N (\eps(\Lambda^{(i)},\Lambda^{(j)})+tg_1(\Lambda^{(i)},\Lambda^{(j)}))}{\prod_{i=1}^N (\eps(\Lambda^{(i)},\Lambda^{(i)})+tg_1(\Lambda^{(i)},\Lambda^{(i)}))}}
\ee
%(we don't write obvious limit here).
 But there is only one solution, so the numerator equals $g(\Lambda^{(1)},\Lambda^{(1)})$ and denominator equals $g(\Lambda^{(1)},\Lambda^{(1)})$. So complanart equals 1. Due to similar reasons, \textbf{ complanart equals 1 always when \\  $\mathbf{N=r_1r_2\cdots r_{n-1}<n}$}. For example, complanart equals 1 for any system of appropriate number of linear equations. 
\subsubsection{Open questions}
The first question is about the choices of $g_1$: (\ref{gone}),(\ref{gtwo}),(\ref{gthree}). These formulas are not $GL_n$-invariant.
% And, despite that complanart does not depend on $g_1$, the higher terms on t depend on the choosing of $g_1$, and if we want to use them, we will have to deal with this $g_1$-dependence.
 The second question: since we do not know appropriate canonical $g_1$, we may take: $g(x,y,\cdots)=\eps(x,y,\cdots)+t_1g_1(x,y,\cdots)+t_2g_2(x,y,\cdots)+\cdots$, and now look at the same limit, but taken on different $t$ variables in different consequence.
 The third question is: by taking a limit at (\ref{sovptwo}-\ref{sovpfour}), only term with the lowest degree of $t$ is considered. It is possible, that the terms of higher degrees contain information about higher degenerations of the system. For example, vanishing of some higher term(s) may correspond to existence of two coinciding roots and so on. This point can be even more interesting considering the higher terms in the case of many $t$.
%  (since, as have been already mentioned, for two vectors to projectively coincide in 3-dimensional space, it is necessary {\it two} conditions - and these conditions may be first term on t=0 (with the smallest degree on t) and second term on t=0 (the term with next higher degree)).
   The theory of higher complanarts would also be very interesting and useful in applications.
\subsubsection{Examples of complanarts: $n=2$, complanarts are discriminants}
\paragraph{Quadratic equation}
Our first example is:
\be
f(x)=a(x_1)^2+bx_1x_2+c(x_2)^2, \nonumber
\ee
or in non-homogeneous variable $z\equiv x_1/x_2$:
\be
 az^2+bz+c=0 \nonumber
\ee
Take $g(x,y)=\eps(x,y)+tg_1(x,y)=x_1y_2-x_2y_1+t(x_1y_1+x_1y_2+x_2y_1+x_2y_2)$. For this equation and this $g_1$:
\be
 \prod_{i,j}g(\Lambda^{(i)},\Lambda^{(j)})=t^2(a+c-b)^2(4ac-b^2+t^2(b^2-2ab+a^2-2bc+2ac+c^2)) \nonumber  \\
 \prod_{i}g(\Lambda^{(i)},\Lambda^{(i)})=t^2(a+c-b)^2 \nonumber
\ee
It is easily seen, that $C=4ac-b^2$, and this expression coincides with discriminant of quadratic polynomial. Now take another $g_1=x_1y_2$. For this $g_1$:
\be
 \prod_{i,j}g(\Lambda^{(i)},\Lambda^{(j)})=t^2ac(4ac-b^2+t(4ac-b^2)+t^2(ac-b^2)) \nonumber \\
 \prod_{i}g(\Lambda^{(i)},\Lambda^{(i)})=t^2ac \nonumber
\ee
The expressions have changed. But the value of complanart is still $4ac-b^2$.
\paragraph{Cubic polynomial}
\be
 f(x)=ax^3+bx^2y+cxy^2+dy^3
\ee
Complanart of this polynomial is equal $4ac^3+4db^3+27a^2d^2-b^2c^2-18abcd$, and it again coincides with the discriminant of the polynomial. Other expressions are very long in this case, and we do not write them here.
\subsubsection{Examples of complanarts: $n=3$ \label{sovpexample}}
\paragraph{Two quadratic polynomials, 1}
\be
 f_1(x)=a_{11}(x_1)^2+a_{21}x_1x_2+a_{13}x_1x_3 \nonumber \\
 f_2(x)=b_{22}(x_2)^2+b_{21}x_2x_1+b_{23}x_2x_3 \nonumber
\ee
%In this example we will illustrate the statement, that you cannot evaluate complanart with arbitrary $g_1$.
Take, for example, $g(x,y,z)=\eps(x,y,z)+tg_1(x,y,z)=(1+t)x_1y_2z_3-x_1y_3z_2+(1+t)x_2y_3z_1-x_2y_1z_3+(1+t)x_3y_1z_2-x_3y_2z_1$. Than all three terms  in (\ref{sovpthree}), namely \\ $\prod\limits_{i,j,k=1}^N g(\Lambda^{(i)},\Lambda^{(j)},\Lambda^{(k)}),\quad\prod\limits_{i=1}^N g(\Lambda^{(i)},\Lambda^{(i)},\Lambda^{(i)}),\quad\prod\limits_{i,j=1}^N g(\Lambda^{(i)},\Lambda^{(i)},\Lambda^{(j)})$, vanish, so the limit cannot be evaluated. Nevertheless, if we take, for example:
\be
 g_1=\sum_{i,j=1}^nx_iy_j=x_1y_1+x_1y_2+x_1y_3+x_2y_1+x_2y_2+x_2y_3+x_3y_1+x_3y_2+x_3y_3, \nonumber
\ee 
i. e. choose $g_1$ according to (\ref{gthreevar}), this problem is eliminated. The limit (\ref{sovpthree}) equals
\be
\lim_{t\rightarrow 0}{\frac{\left(\prod_{i,j,k=1}^N g(\Lambda^{(i)},\Lambda^{(j)},\Lambda^{(k)})\right)\left(\prod_{i=1}^N g(\Lambda^{(i)},\Lambda^{(i)},\Lambda^{(i)})\right)^2}{\left(\prod_{i,j=1}^N g(\Lambda^{(i)},\Lambda^{(i)},\Lambda^{(j)})\right)^3}}= \nonumber  \\
 (a_{13})^{12}(b_{23})^{12}(a_{21}b_{23}-a_{13}b_{22})^6(b_{23}a_{11}-a_{13}b_{21})^6(-b_{22}a_{11}a_{13}b_{23}-b_{21}a_{21}a_{13}b_{23}+b_{21}a_{13}^2b_{22}+b_{23}^2a_{21}a_{11})^6 \nonumber
\ee
Complanart can be calculated also by bare hands, i. e. by solving equations $f_1(x)=0,f_2(x)=0$ and then substituting these solutions in (\ref{sovpthreevar}).
% NORMALIZATION - nado otkomentit
%\footnote{Of course, complanart depends also {\it on the normalization} of roots. So all the equalities are written in some nornalization}.
Calculated complanart equals:
\be
 (a_{13})^{4}(b_{23})^{4}(a_{21}b_{23}-a_{13}b_{22})^2(b_{23}a_{11}-a_{13}b_{21})^2(-b_{22}a_{11}a_{13}b_{23}-b_{21}a_{21}a_{13}b_{23}+b_{21}a_{13}^2b_{22}+b_{23}^2a_{21}a_{11})^2 \nonumber
\ee
Thus, the formula (\ref{sovpthree}) holds.
%Thus, complanart is given by formula (\ref{sovpthree}) correctly.
\paragraph{Two quadratic polynomials, 2 \label{sovpab}}
This example is just the particular case of previous one, but it is of great importance for further considerations  of sect  \ref{example}:
\be
%f_1(x)=(x_1)^2+2ax_1x_2-x_1x_3 \\
%f_2(x)=(x_2)^2+2bx_2x_1-x_2x_3
f_1(x)=x^2+2axy-xz \\
f_2(x)=y^2+2bxy-yz
\ee
The complanart cubed, and the limit in the formula (\ref{sovpthree}) yield:
\be
C^3=(1-2a)^{12}(1-2b)^{12}
\ee
\section{Eigenvectors, eigenvalues and $GL_n$ canonical representation of non-linear maps \label{eigenvectors}}
\subsection{Non-linear eigenvectors, eigenvalues and how they can be found}
Non-zero vector $z_i$ is called eigenvector of $A$, if it satisfies (with some $\lambda$):
\be
A_i^{i_1i_2\cdots i_s}z_{i_1}z_{i_2}z_{i_3}\cdots z_{i_s}=\lambda(z)z_i \label{eigvect}
\ee
A polynomial $\lambda(z)$ of degree $s-1$ is called eigenvalue. Eigenvalue is not a number, because the homogeneity constraint is imposed. The homogeneity constraint is nothing but requirement that any eigenvector can be multiplied by some non-zero number and remain eigenvector. There is also a non-linear analogue of characteristic polynomial of a map, namely:
% Why $\lambda$ is a polynomial now (unlike linear case)?
%$\lambda(z)$ is polynomial in order to save homogeneity, i. e. that all components of eigenvector can be multiplied by a non-zero number and it remained eigenvector. If we want to find eigenvectors/eigenvalues, analogy with linear algebra case supposes us to calculate the characteristic polynomial:
\be
 Ch_A\equiv R\{A_i(z)-\lambda(z)z_i\} \label{charmnogtwo}
\ee
By definition of resultant, $Ch_A$ turns to zero iff exists an eigenvector, corresponding to the polynomial $\lambda(z)$. The optimal way to find eigenvectors/eigenvalues in non-linear algebra differs from the way in linear algebra. Solving equation $Ch_A=0$ with respect to $\lambda$ is complicated, because it is a non-homogeneous equation of $M_{n|s-1}=\frac{(n+s-2)!}{(n-1)!(s-1)!}$  coefficients of $\lambda$ of the degree $c_{n|s}=\frac{s^n-1}{s-1}$. It turns out, that $Ch_A$ is not an arbitrary polynomial, it has a structure - namely, this polynomial is always decomposable on {\it linear} in the coefficients of $\lambda$ factors. It is much simpler to find these factors at first, and then find eigenvalues by solving linear equations. To decompose $Ch_A$, it is necessary to find eigenvectors, this is described in sect \ref{decomp}. The method of eliminating $\lambda$ from the equation (\ref{eigvect}) is described in sect \ref{unitaryeig}.
\subsection{Zero, nonzero and unitary eigenvectors \label{unitaryeig}}
One encounters two cases considering a particular eigenvector $z_i$: $\lambda(z)\ne 0$ and $\lambda(z)=0$. Consider the first case. All eigenvectors with $\lambda(z)\ne 0$ we call non-zero eigenvectors. This notation can not cause any misinterpretation, because each eigenvector is by definition non-zero in the sense that not all its components are zeros. Non-zero eigenvector can be rescaled by any non-zero number, because the system of equation (\ref{eigvect}) is homogeneous system. For example, $z_i$ can be rescaled as follows:  $y_i=\frac{z_i}{\lambda(z)^{1/(s-1)}}$. Now, $\lambda(y)=\frac{\lambda(z)}{(\lambda(z))^{\frac{s-1}{s-1}}}=1$, and 
\be
A_i(y)=y_i \label{oneeig}
\ee
The eigenvectors, obeying (\ref{oneeig}), are called unitary eigenvectors. $\lambda$ was eliminated from  equation for eigenvectors/eigenvalues. Thus, all non-zero eigenvectors
% (and unitary eigenvectors being a particular case of nonzero eigenvectors).
can be rescaled to unitary vectors, which satisfy (\ref{oneeig}). \\
 Now consider the case $\lambda(z)=0$. Such eigenvectors are called zero-eigenvectors. Any zero eigenvector satisfies the equation:
\be
A_i(z)=0, \label{zeroeig}
\ee
which also does not contain $\lambda$. \\
% So, the direct statement was proven \prop{All eigenvectors of $A$ can be found among the solutions of (\ref{oneeig}) or (\ref{zeroeig}), and, otherwise, any non-zero solution of (\ref{oneeig}) or (\ref{zeroeig}) is eigenvector of $A$ \label{ridlambda}}
%The inverse statement is that 
Any solution of (\ref{oneeig}) or (\ref{zeroeig}) is an eigenvector. To prove this, it is sufficient to construct $\lambda$ satisfying the equation (\ref{eigvect}) with this $z_i$.
%The inverse statement is proved as follows:
 If $y_i$ is the solution of (\ref{oneeig}), $y_i$ is the eigenvector with any $\lambda$ such that:
\be
 \lambda(y_i)=1, \label{lambdaone}
\ee
as it is easily seen from (\ref{eigvect}). If $z_i$ is the solution of (\ref{zeroeig}), it is eigenvector with any $\lambda$ which satisfies:
\be
 \lambda(z_i)=0 \label{lambdazero}
\ee
Thus, the following statement was proven:
\prop{All eigenvectors of $A$ can be found among the solutions of (\ref{oneeig}) or (\ref{zeroeig}), and, otherwise, any non-zero solution of (\ref{oneeig}) or (\ref{zeroeig}) is eigenvector of $A$ \label{ridlambda}}
%We see, that there are two kinds of eigenvectors. Eigenvectors of first kind obey (\ref{eigvect}) with $\lambda(z)\ne 0$. We will call them {\it non-zero eigenvector}. As we have seen just now, having a non-zero eigenvector, we can renozmalize it to obey the equation (\ref{oneeig}). Such normalized eigenvector we will call {\it unitary eigenvector}. All unitary eigenvectors obey (\ref{lambdaone}) {\it with corresponding $\lambda$}. And the eigenvectors of the second kind are eigenvectors, obeying (\ref{zeroeig}). We will call them "zero" eigenvectors (it can't cause misinterpretaion, see the footnote on this page). 
So, the easy method of finding eigenvector/eigenvalues is: solve equations (\ref{oneeig}-\ref{zeroeig}) to find eigenvectors, and then find $\lambda$ by algorithm, described above. 
%In the "standart" approach through charactiristical polynomial we would solve one equation of high\footnote{and we will determine this degree} degree of many\footnote{$M_{n|s-1}=\frac{(n+s-2)!}{(n-1)!(s-1)!}$, this is a number of independent coefficients of polynomial of $n$ variables of degree $s-1$, in this case $\lambda$} variables $Ch_A(\lambda)=0$, then solve {\it many} systems $A_i(z)-\lambda(z)z_i=0$ of $n$ equations for $n$ variables (for each $\lambda$ we find). In our method we have to solve only two systems of $n$ equations of $n$ variables (and then some number of linear systems to determine $\lambda$). The sense, in which eigenvalues "does not matter" (which was announced in \ref{introdeig}), consists of Prop.\ref{ridlambda}: you can at first find all eigenvectors and then elementary find $\lambda$s. 
\subsection{Decomposability of characteristic polynomial \label{decomp}}
Since to find $\lambda$ one has to solve {\it linear} on $\lambda$ equations (see previous subsection), the set of $\lambda$, which are eigenvalues of $A$, is a union of planes in the space of all $\lambda$. Other way to reformulate this statement is to say: the characteristic polynomial $Ch_A(\lambda)$ is decomposable on {\it linear} on $\lambda$ factors. The condition of turning to zero one of the factors of $Ch_A$ defines the plane in the space of all $\lambda$. These factors are the following:
% And now we can say, which are these factors:
\be
 R\{A_i(z)-\lambda(z)z_i\}=M(A)\prod_{\mu=1}^{n_1}(1-\lambda(e^{(\mu)}))\prod_{\nu=1}^{n_2}(\lambda(z^{(\nu)})), \label{razlchar}
\ee
where $e^{(\mu)}$ is the $\mu$-th unitary eigenvector, and $n_1$ is the number of them, $z^{(\nu)}$ is $\nu$-th zero eigenvector, and $n_2$ is the number of them, and $M(A)$ depends only on coefficients of $A$, not on the coefficients of $\lambda$. Indeed, all possible eigenvectors are found among the solutions (\ref{oneeig}-\ref{zeroeig}) modulo projective rescaling.
% Therefore all possible eigenvalues satisfy either the equation (\ref{lambdaone}) with one of the unitary eigenvectors, and in this case $\lambda$ turns to zero (\ref{razlchar}) due to one of the terms $(1-\lambda(e_{\mu}))$, or the equation (\ref{lambdazero}) with one of the zero eigenvectors, and in this case $\lambda$ turns to zero (\ref{razlchar}) due to one of the factors $\lambda(z_{\nu})$. 
Therefore all possible eigenvalues satisfy either the equation (\ref{lambdaone}) with one of the unitary eigenvectors, either the equation (\ref{lambdazero}) with one of the zero eigenvectors. If $\lambda$ satisfies (\ref{lambdaone}), r. h. s. of (\ref{razlchar}) turns to zero due to one of the factors $(1-\lambda(e_{\mu}))$. If $\lambda$ satisfies (\ref{lambdazero}), r. h. s. (\ref{razlchar}) turns to zero due to one of the factors $\lambda(z_{\nu})$. 
\subsection{Number of eigenvectors \label{numeig}}
The systems of equations (\ref{oneeig}) and (\ref{zeroeig}) can be merged to one system of homogeneous equations by adding an auxiliary variable $y$:
\be
A_i(x)=(y)^{s-1}x_i \label{oneeigodn}
\ee
(\ref{oneeigodn}) is a system of $n$ homogeneous equations of $n+1$ variables $x_1,x_2,\cdots,x_n,y$. If a solution of (\ref{oneeigodn}) has $y\ne 0$, we can consider the vector $x_1/y,\cdots,x_n/y,1$, which is a solution both of (\ref{oneeigodn}) and of (\ref{oneeig}), and then first $n$ components of it will be an unitary eigenvector. If a solution of (\ref{oneeigodn}) has $y=0$, the first $n$ components $x_1,\dots,x_n$ of this solution are simultaneously a solution of (\ref{zeroeig}) and vector of them is a zero eigenvector. The examples of finding eigenvectors using (\ref{oneeigodn}) are presented in sect \ref{canrepres} and in sect  \ref{example}. So, the following statement holds:
\prop{All eigenvectors of $A$ can be found among the solutions of (\ref{oneeigodn}) (these solutions should be normalized, if necessary).}
Consider non-degenerate maps, $R\{A_i(z)\}\ne 0$. Firstly, for such maps the equation (\ref{zeroeig}) has no solutions with at least one component being non-zero. The condition $R\{A_i(z)\}\ne 0$ is by definition of resultant the condition of non-existence of non-trivial solutions of (\ref{zeroeig}). That is, non-degenerate map does not have zero eigenvectors. The same statement holds in the linear algebra. This means, that $n_2=0$ and in (\ref{razlchar}) there is no factor $\prod\limits_{\nu=1}^{n_2}\lambda(z_{\nu})$,
\be
 R\{A_i(z)-\lambda(z)z_i\}=M(A)\prod_{\mu=1}^{n_1}(1-\lambda(e_{\mu})) \label{razlnondegcharM}
\ee
Setting in (\ref{razlnondegcharM}) $\lambda=0$, one obtains $M(A)=R\{A\}$. So
\be
 R\{A_i(z)-\lambda(z)z_i\}=R\{A\}\prod_{\mu=1}^{n_1}(1-\lambda(e_{\mu})) \label{razlnondegchar}
\ee
For non-degenerate maps, $n_1$ can be easily determined.
%Now we can determine $n_1$ (under $R\{A\}\ne 0$).
 Consider again (\ref{oneeigodn}). If $R\{A\}\ne 0$, it has no  non-zero solutions with $y=0$ (the existence of such solutions would imply the existence of non-zero solutions of (\ref{zeroeig}), but this is prohibited by $R\{A\}\ne 0$). As a system of $n$ homogeneous equations (each of degree $s$,$\quad s$ - a degree of $A$) of $n+1$ variables, in general case (\ref{oneeigodn}) has $s^n$ projectively non-equivalent solutions, see \cite{Galois}. 
%(for an example of such counting see \ref{Vieta}).
 It has one solution which is not eigenvector:
\bs{r}
 x_1=x_2=\cdots=x_n=0 \label{neint} \\
 y=1  % \label{neinttwo}
\es
The number of solutions of (\ref{oneeigodn}) with at least one $x$-component being non-zero is $s^n-1$.
% So the number of potential eigenvectors is now $s^n-1$, but we know, that not all $x_1,\cdots,x_n$ equal to $0$ (otherwise it would be projectively equivalent to (\ref{neintone}-\ref{neinttwo})) and $y\ne 0$.
  Some of these solutions yield projectively equivalent eigenvectors (in the space $x_1,\cdots,x_n$). Let $x^{(0)}_1,x^{(0)}_2,\cdots,x^{(0)}_n,y^{(0)}$ is a solution of (\ref{oneeigodn}). Then $x^{(0)}_1,x^{(0)}_2,\cdots,x^{(0)}_n,\omega_{s-1}y^{(0)}$, where $\omega_{s-1}$ is a root of degree $s-1$ from 1, is also solution of (\ref{oneeigodn}). Since in complex plane there is $s-1$ distinct roots of 1, $s-1$ projectively inequivalent (in the space $x_1,\cdots,x_n,y$) solutions of (\ref{oneeigodn}) lead to one eigenvector (or in projective equivalent vectors in $x_1,\cdots,x_n$). So the number of eigenvectors in general case is $c_{n|s}=\frac{s^n-1}{s-1}$. This formula was previously obtained in \cite{INT}, but was obtained from considerations for diagonal maps. So, characteristic polynomial has degree $c_{n|s}$ on the coefficients of $\lambda$.
%But (\ref{oneeigodn}) has a $s^n$ solutions only in general case, and what are the cases besides the "general case"?
 Some of $s^n$ solutions of (\ref{oneeigodn}) can coincide (this phenomenon is controlled by complanart and higher complanarts, see \ref{subsovpad}). In this case there will be less eigenvectors than $c_{n|s}$.
% At the other side, the range of entire system (\ref{oneeigodn}) can decrease, and it may have not $s^n$ projectively-nonequivalent solutions, but one-, or two-dimensional (or more-dimensional) space of projectively non-equivalent solutions.
%  I, however, have a feeling, that 
It should be emphasized one more time, that all counting, leaded to the formula $c_{n|s}=\frac{s^n-1}{s-1}$, was carried out under the condition $R\{A\}\ne 0$. For example, for unit maps (s. \ref{unitmaps}) all the vectors in the space are eigenvectors.
% The situation of degenerate maps is poorly investigated. 
\subsection{$GL_n$ canonical representation of non-linear map \label{canrepres}} 
In linear algebra linear map (and quadratic form) have canonical representation (i. e., become diagonal) in the basis of its eigenvectors. The situation in non-linear algebra is similar, but there are some differences. Firstly, in general case non-linear map cannot be brought to diagonal from, because the group $GL_n$ has $n^2$ parameters, and non-linear map has more parameters. The second difference is, that non-linear map in general case has $c_{n|s}$ eigenvectors, see \ref{numeig}. The number of eigenvectors is greater than $n$, the dimension of the space. So there is uncertainty in canonical representation of non-linear map, consisting in the freedom  to choose one set of $n$ eigenvectors from $c_{n|s}$ eigenvectors to be basis of the space. This uncertainty is yet poorly studied.
% Nevertheless, let us see what is the canonical representation of non-linear map. 
To obtain canonical representation of non-linear map, one should choose as basis vectors any $n$ linear independent zero or unitary eigenvectors of the map. The eigenvectors of the system can be found by solving the system of equations (\ref{oneeigodn}). %After eigenvectors have been found, one can choose $n$ of eigenvectors to be a basis vectors.
 Non-zero eigenvectors should be normalized to make them unitary eigenvectors. 
% Each of these vectors should be either zero eigenvector or unitary eigenvector, 
%All non-zero eigenvectors should be normalized to unitary eigen
Consider first what happens with the components of the map, corresponding to unitary eigenvector. If unitary eigenvector $z_i$ is chosen to be a $j$-th vector of basis, $z_i=\delta_i^j$.
%, it has all coordinates equal to zero, except for $j$-th coordinate, which equals $1$.
 By definition of unitary eigenvector, $A_i^{i_1i_2\cdots i_s}z_{i_1}z_{i_2}z_{i_3}\cdots z_{i_s}=z_i$. Substituting $z_i=\delta_i^j$, we obtain $A_i^{jj\cdots j}=\de^j_i$. If zero eigenvector $y_i$ is chosen to be $k$-th vector of basis, then $y_i=\delta_i^k$. The equation for zero eigenvectors is: $A_i^{i_1i_2\cdots i_s}y_{i_1}y_{i_2}y_{i_3}\cdots y_{i_s}=0$. and therefore $A_i^{kk\cdots k}=0$. Summarizing the statements about the canonical form:
\be 
%\boxed{
A_i^{jj\cdots j}=\de^j_i \quad {\rm If } \quad j \quad { \rm corresponds \quad to\quad unitary\quad eigenvector\quad in \quad basis}  \\
A_i^{kk\cdots k}=0 \quad{\rm If \quad} k {\quad \rm corresponds\quad to\quad zero\quad eigenvector\quad in\quad basis} 
%}
\ee 
By this method, only the components with all upper indices equal are fixed, total $n^2$ components. More components can not be fixed, since there are only $n^2$ free parameters in $GL_n$ group. The rest components of $A$ remain arbitrary numbers. It is these numbers that determine the map in the non-linear case. In linear case the map is defined by $n$ eigenvectors and $n$ eigenvalues. In non-linear case the map is defined by $n$ of its eigenvectors (not all!) and by these numbers, namely by the components with not all upper indices equal. One more difference in these representations is that in non-linear case the normalization of non-zero eigenvectors is important, but in linear case it is not. 
\subsubsection{Example of canonical representation \label{excanrepres} }
%All the statements about canonical representation of the map are illustrated on one example. 
The example is the bringing the quadratic map of two variables to canonical representation:
\be
\left(
\ba{c}
x_1 \\
x_2 
\ea
\right) \rightarrow \left(
\ba{c}
A_1^{11}(x_1)^2+2A_1^{12}x_1x_2+A_1^{22}(x_2)^2\\
A_2^{11}(x_1)^2+2A_2^{12}x_1x_2+A_2^{22}(x_2)^2
\ea
\right) \label{examplequadtwo}
\ee
To obtain canonical representation of non-linear map, one should choose as basis vectors any $n$ linear independent zero or unitary eigenvectors of the map. As it was established in sect \ref{numeig} and \ref{unitaryeig}, to obtain the eigenvectors of the map (\ref{examplequadtwo}), one should solve such systems of equations:
\be
\left\{
\ba{c}
 A_1^{11}(x_1)^2+2A_1^{12}x_1x_2+A_1^{22}(x_2)^2=yx_1 \label{examplesone} \\
 A_2^{11}(x_1)^2+2A_2^{12}x_1x_2+A_2^{22}(x_2)^2=yx_2 %\label{examplestwo}
\ea
\right.
\ee
This is a system of equations (\ref{oneeigodn}) for (\ref{examplequadtwo}), additional homogenizing variable is also called $y$. Any solution of the system (\ref{examplesone}),  $(x_1,x_2,y)$ with $y\ne 0$ is non-zero eigenvector for the map (\ref{examplequadtwo}). Such solution can be made unitary eigenvector by dividing all the components of it by $y$. Then the solution will be $(x_1/y,x_2/y,1)$ and $x_1/y,x_2/y$ are two components of unitary eigenvector. If there is a solution of (\ref{examplesone}) with third component equal $0$, e. g. $(x_1,x_2,0)$, then $(x_1,x_2)$ are the components of zero eigenvector. In some cases, the map can be represented in several forms of (\ref{canvidtwounit}),(\ref{canvidunitzero}),(\ref{canvidtwozero}). For example, if the map has two unitary eigenvectors and one zero eigenvector, it can be represented either in form (\ref{canvidtwounit}) or in the form (\ref{canvidunitzero}).
\paragraph{Two unitary eigenvectors are chosen to be basis}
This choice fixes the components to the following values: $A_1^{11}=1,\quad A_1^{22}=0,\quad A_2^{11}=0, \quad A_2^{22}=1$. Canonical representation of the map in this case:
\be
\left(
\ba{c}
x_1 \\
x_2 
\ea
\right) \rightarrow \left(
\ba{c}
(x_1)^2+2A_1^{12}x_1x_2\\
(x_2)^2+2A_2^{12}x_1x_2
\ea
\right) \label{canvidtwounit}
\ee
%When resultant of the map (\ref{canvidtwounit}), namely $1-4A_1^{12}A_2^{12}$, vanishes, this map has also zero eigenvector $(1-2A_1^{12},1-2A_2^{12})$, and can be represented in the form (\ref{canvidunitzero}).
%The map (\ref{canvidtwounit}) sometimes has also a zero eigenvector, so it can be represented
\paragraph{One unitary and one zero eigenvector are chosen to be basis vectors }
In this case "canonical basis" consists of one zero eigenvector and one unitary eigenvector. Let unitary eigenvector be the first vector of the basis, and zero eigenvector the second. This choice fixes $A^{11}_1=1,\quad A^{22}_1=0, \quad A^{11}_2=0,\quad A^{22}_2=0$. Canonical representation of the map is:
\be
\left(
\ba{c}
x_1 \\
x_2 
\ea
\right) \rightarrow \left(
\ba{c}
(x_1)^2+2A_1^{12}x_1x_2\\
2A_2^{12}x_1x_2
\ea
\right) \label{canvidunitzero}
\ee
%Under some values of $A_1^{12}$ and $A_2^{12}$.
\paragraph{Two zero eigenvectors are chosen to be basis vectors}
In this case "canonical basis" consists of two zero eigenvectors. This choice fixes $A^{11}_1=0,\quad A^{22}_1=0, \quad A^{11}_2=0,\quad A^{22}_2=0$. Canonical representation of the map is:
\be
\left(
\ba{c}
x_1 \\
x_2 
\ea
\right) \rightarrow \left(
\ba{c}
2A_1^{12}x_1x_2\\
2A_2^{12}x_1x_2
\ea
\right) \label{canvidtwozero}
\ee
\subsection{Application of complanart to the theory of eigenvectors \label{sovpappl}}
As it was explained in \ref{decomp}, all the eigenvectors of $A$ can be found by solving this system of equations:
\be
A_i(x)=(y)^{s-1}x_i \label{eigodn}
\ee
This is a system of $n$ homogeneous equations of $n+1$ variables, so it possesses complanart (see \ref{subsovpad}).
% When complanart of this system vanishes, there are complanar solutions of this system.
 Vanishing of complanart of this system means, that there are $n+1$ complanar vectors in the space $(x_1,x_2,\dots,x_n,y)$, i. e. there exists set of $n+1$ indices $j_1,\dots,j_{n+1}$ with no pair of equal indices, such as:
\be
 \eps^{i_1\dots i_{n+1}}\Lambda_{i_1}^{(j_1)}\Lambda_{i_2}^{(j_2)}\dots\Lambda_{i_{n+1}}^{(j_{n+1})}=0 \label{epslambda}
\ee
Here $\Lambda^{(j)}$ stands for $j$-th solution of (\ref{eigodn}), $\Lambda^{(j)}_i$ is $i$-th component of $\Lambda^{(j)}$. Here $i$ runs from $1$ to $n+1$, $\Lambda^{(j)}_{n+1}$ is $y$-component of $j$-th solution. Let one of $j$ indices, for example, $j_1$, corresponds to the solution of (\ref{eigodn}) of type (\ref{neint}), namely $(x_1=0,x_2=0,\dots,x_n=0,y=1)$. Then the sum over $i_1$ in (\ref{epslambda}) is reduced to one term: 
\be
\eps^{(n+1)i_2\dots i_{n+1}}\Lambda_{n+1}^{j_1}\Lambda_{i_2}^{(j_2)} \dots \Lambda_{i_{n+1}}^{(j_{n+1})} =\eps^{(n+1)i_2\dots i_{n+1}}\Lambda_{i_2}^{(j_2)}\dots\Lambda_{i_{n+1}}^{(j_{n+1})} = \nonumber \\
 \pm \sum_{i_2,\dots,i_{n+1}=1}^{n}\eps^{i_2\dots i_{n+1}}\Lambda_{i_2}^{(j_2)}\dots\Lambda_{i_{n+1}}^{(j_{n+1})} \label{epsinit}
\ee
The equality $\Lambda_{n+1}^{j_1}=1$, following from (\ref{neint}), was used. The explicit symbol of the sum in last line is to emphasize, that $\eps$ now is $n$-dimensional, not $n+1$-dimensional and the sum over indices goes to $n$, not to $n+1$, as in the first line. Expression (\ref{epsinit}) is up to sign a condition of $\Lambda_{i_2}^{(j_2)},\dots,\Lambda_{i_{n+1}}^{(j_{n+1})}$ to be complanar {\it  in $n$-dimensional space of initial coordinates $x_1,x_2,\dots,x_n$}. If the solution (\ref{neint}) does not enter in the formula (\ref{epslambda}), the formula (\ref{epslambda}) does not possess such a simple interpretation. So, \textbf{vanishing of complanart of the system (\ref{eigodn}) means either that there are $n$ eigenvectors, complanar in usual space $x_1,\dots,x_n$, or that there are $n+1$ eigenvectors, complanar in the space with additional homogenizing variable, in space $ \mathbf{(x_1,\dots,x_n,y)}$.} \\
%, but it doesn't mean that there are $n$ complanar vectors in the space of $(x_1,x_2,\dots,x_n)$.
%Consider what is formula (\ref{epslambda}) when the solution (\ref{neint}) does not present in it.
\subsubsection{Example}
%Consider one example of the above statements.
 This example is consideration of non-degenerate map of two variables, $n=2$ and $R\{A\}\ne 0$. Since $R\{A\}\ne 0$,the map has only non-zero eigenvectors, which can be rescaled to be unitary eigenvectors. Let three unitary eigenvectors $(x_1,x_2,1),\quad (u_1,u_2,1),\quad (z_1,z_2,1)$ enter to the formula (\ref{epslambda}). Then formula (\ref{epslambda})  becomes:
\be
 x_1u_2z_3-x_1u_3z_2+x_2u_3z_1-x_2u_1z_3+x_3u_1z_2-x_3u_2z_1=x_1u_2-x_2u_1+x_2z_1-x_1z_2+u_1z_2-u_2z_1=0 \label{threesloj}
\ee
The solution of type (\ref{neint}) equals in this case: $(0,0,1)$. Let $(x_1,x_2,1),(z_1,z_2,1)$ and the solution of type (\ref{neint}) enter to the formula (\ref{epslambda}). Then the formula (\ref{epslambda}) becomes:
\be
 x_1z_2-x_2z_1=0, \label{threeprost}
\ee
or simply condition of complanarity of two {\it two-dimensional} vectors $(x_1,x_2)$ and $(y_1,y_2)$ in {\it two-dimensional space}. Vanishing of complanart of system (\ref{eigodn})  means, that there are such eigenvectors $(x_1,x_2,1),\quad (u_1,u_2,1),\quad (z_1,z_2,1)$, that either (\ref{threesloj}) or (\ref{threeprost}) holds. For degenerate map only the equality (\ref{threesloj}) would change. 
% Vanishing of complanart of the system (\ref{eigodn}) means, that there are three eigenvectors (these vectors are normalized to be unitary eigenvectors), such as $x_1u_2z_3-x_1u_3z_2+x_2u_3z_1-x_2u_1z_3+x_3u_1z_2-x_3u_2z_1=x_1u_2-x_2u_1+x_2z_1-x_1z_2+u_1z_2-u_2z_1=0$.
% Since in this example $R\{A\}\ne 0$, all eigenvectors are non-zero, and can be renormalized to be unitary eigenvectors. % In this example $(x_1,x_2,1),\quad (y_1,y_2,1),\quad (z_1,z_2,1)$ are unitary eigenvectors, and therefore $x_3=y_3=z_3=1$.
% The solutions are complanar in the space with an additional variable, namely in the space $(x_1,x_2,\dots,x_n,y)$.
%  So  does not mean that there are $n$ complanar vectors in the space of $(x_1,x_2,\dots,x_n)$, it 
One more example of using complanart to determine degeneracy of eigenvector pattern is given in sect \ref{exampleig}.
\subsection{How to exclude $\lambda$ by one more method \label{everyvecteig}}
In sect \ref{unitaryeig} the method of excluding $\lambda$ from equations for eigenvectors by renormalizing vectors was explained. There is one more method of excluding $\lambda$. 
%From this method the condition for a map to be a unit map can be extracted. 
This method was mentioned in \cite{INT}.
The method is based on following statement:
\prop{All the eigenvectors of the map $A$ can be found among the solutions of system of equations:
\be
A_i(z)z_j-z_iA_j(z)=0, \label{ohnelambda}
\ee
and otherwise, any solution of (\ref{ohnelambda}) is an eigenvector.
}
\textbf{Proof} If $z_i$ is an eigenvector of $A_i, \quad A_i(z)=\lambda(z)z_i$. Substituting this equality to (\ref{ohnelambda}) yields: $A_i(z)z_j-A_j(z)z_i=\lambda(z)z_iz_j-\lambda(z)z_jz_i=0$. Otherwise, let $z_i$ is the solution of (\ref{ohnelambda}). Multiply both sides of equation (\ref{ohnelambda}) by a vector $v^i$, orthogonal to $z_i$. Then one obtains $z_jA_i(z)v^i=A_j(z)z_iv^i=0$. So, $A_i(z)$ is orthogonal to any vector, orthogonal to $z_i$, therefore $A_i(z)=C\cdot z_i$. One can obtain $\lambda(z)$, corresponding to this eigenvector by solving the equation $\lambda(z)=C$. \\
At first glance, in (\ref{ohnelambda}) there are $\frac{n(n-1)}{2}$ non-trivial equations. But only $n-1$ of them are independent: you can fix $i$ and get $n-1$ equations for $j\ne i$, and all other equations will follow from these. This fact was previuosly noted in \cite{INT}, but without a proof. This is a proposition \ref{Allfromfixj} of sect. \ref{eigstat}, and it is proven there. \\
 This is one more method to show linear decomposability of characteristic polynomial (\ref{charmnogtwo}); one again has to solve {\it linear} on $\lambda$ equations. For example, for (\ref{examplequadtwo}) the system of equations (\ref{ohnelambda}) reduces to single equation:
\be 
A_2^{11}(x_1)^3+(2A_2^{12}-A_1^{11})x_2(x_1)^2+(A_2^{22}-2A_1^{12})(x_2)^2x_1-A_1^{22}(x_2)^3=0 \label{condunit}
\ee
\subsubsection{Unit maps \label{unitmaps}} 
Using the system of equations (\ref{ohnelambda}), one can establish the condition of each vector of the space to be an eigenvector of the map. This condition is condition of vanishing of all the coefficients in the equations (\ref{ohnelambda}). The map, for which each vector of the space is an eigenvector, is called unit map. Any unit map can be represented as $A_i(z)=\mu(z)z_i$, where $\mu(z)$ is a homogeneous polynomial of degree $s-1$. Characteristic polynomial for unit maps is identically zero, $R\{A^i(z)-\lambda(z)z^i\}\equiv 0$. Substituting $A_i(z)=\mu(z)z_i$, one obtains: $R\{A^i(z)-\lambda(z)z^i\}=R\{\mu(z)z_i-\lambda(z)z^i\}=R\{(\mu(z)-\lambda(z))z_i\}=R\{\lambda'(z)z_i\}$. The system of equations $\lambda'(z)z_i=0$ always has non-trivial solution since $\lambda'(z)=0$ always has non-trivial solution, because it is one algebraic equation on $n$ variables. So, $R\{\lambda'(z)z_i\}=0$ and $R\{A^i(z)-\lambda(z)z^i\}\equiv 0$. Unit map maps each vector of the space proportional by itself. This is a generalization of the notion of unit map in non-linear algebra. For example, the condition of the map (\ref{examplesone}) to be unit is:
\be
\left\{
\ba{r}
A_2^{11}=0 \\
2A_2^{12}-A_1^{11}=0 \\
A_2^{22}-2A_1^{12}=0 \\
A_1^{22}=0
\ea
\right.
\ee
These conditions are just the conditions of vanishing of coefficients in (\ref{condunit}).
% One can establish the condition of every vector to be an eigenvector. This condition is vanishing of all the coefficients in the equations (\ref{ohnelambda}).
% These equations are also discussed in s.\ref{unitmaps}.
% In this form it is easily seen, that if all the coefficients of (\ref{ohnelambda}) will be zero, any vector would be eigenvector of the map (more precisely, each vector of the space is zero eigenvector or non-zero eigenvector, and in this case it is proportional to an unitary eigenvector). If all the coefficients in these polynomials will be equal to zero, this map will be a unit map.
\section{Polynomial differential equations \label{sectdiffur}}
\subsection{Eigenvectors as stationary points \label{eigstat}}
%Now we are going to discuss such differential equations:
As it was mentioned in introduction, when one considers the stability of the stationary point of a system of differential equations, one encounters such systems of equations:
\be
 \dot{x}_i=A_i^{i_1i_2\cdots i_s}x_{i_1}x_{i_2}x_{i_3}\cdots x_{i_s} \label{diffurtwo}
\ee
It is well known, that if the degree of the map equals 1, the solution of this equations is expressed through eigenvectors and eigenvalues of linear map $A_i^j$. In the non-linear case, eigenvectors also play an important role in solving these equations, for example, they entirely determine the phase diagram of the system. We begin by rewriting these equations in non-homogeneous variables. For doing this, make the change of "time" variable $dt'=\mu(x)dt$, where $\mu(x)$ is a homogeneous polynomial of degree $s-1$. Now the system is:
\be
 \frac{\p x_i}{\p t'}=\frac{A_i(x)}{\mu(x)} \label{homdiffur}
\ee
Both the sides of equation (\ref{homdiffur}) are homogeneous expressions on $x$ with the same degree of homogeneity, i. e. if $x$ is scaled  $x'=kx$, both the sides of equation will be scaled equally\footnote{This explains the choose of degree of $\mu$}. Denoting $\xi_i\equiv\frac{x_i}{x_1}$:
\be
\frac{\p \xi_i}{\p t'}=\frac{\dot{x}_ix_1-\dot{x}_1x_i}{(x_1)^2}=\frac{A_i(x)x_1-A_1(x)x_i}{\mu(x)(x_1)^2}=\frac{A_i(\xi)-A_1(\xi)\xi_i}{\mu(\xi)} \label{neodndiffur}
\ee
 The equalities $\xi_1\equiv 1$ and $\dot{\xi}_1\equiv 0$ were used. In s.\ref{everyvecteig} is proven a statement, that a vector $x_i$ satisfies:
%It is obvoius from this representation, that eigenvectors of $A$ are the only stationary points of the system \ref{neodndiffur}. Indeed, see \ref{everyvecteig}, there it is explained, that eigenvectors and only they obey 
\be
A_i(x)x_j-x_iA_j(x)=0 \label{ohnelambdatwo}
\ee
iff it is an eigenvector of $A$. % From (\ref{neodndiffur}) it is easily seen, that
Stationary points of (\ref{neodndiffur}) obey the equations:
% But in \ref{ohnelambdatwo} there are $\frac{n(n-1)}{2}$ equations, and 
\be
A_i(x)x_1-A_1(x)x_i=0 \label{ohnelambaone}
\ee
These equations are the equations (\ref{ohnelambdatwo}) with fixed $j=1$.
%This system of equations is the part of the system (\ref{ohnelambdatwo}), 
%are only $n-1$ (if $i=1$, this equation will identically be zero). 
%It turns out (firstly was stated in \cite{INT}), that
\prop{All $\frac{n(n-1)}{2}$ equations (\ref{ohnelambdatwo}) follow from $n-1$ equations (\ref{ohnelambdatwo}) with some fixed $j$.\label{Allfromfixj}}
The system of equations (\ref{ohnelambaone}) is the system (\ref{ohnelambdatwo}) with fixed $j=1$. For any other $j$ the proof is similar. Let $x$ obey (\ref{ohnelambaone}). One can consider the solution of the system (\ref{homdiffur}), which at $t=0$ coincides with $x$. Then for any i $\frac{\p }{\p t'}\left(\frac{x_i}{x_1}\right)=0$, see (\ref{neodndiffur}). So for any i,j:$\frac{\p }{\p t'}\left(\frac{x_i}{x_j}\right)=\frac{\p }{\p t'}\left(\frac{x_i}{x_1}/\frac{x_j}{x_1}\right)=0$, and then, looking again at (\ref{neodndiffur}), $A_i(x)x_j-x_iA_j(x)=0$.
% If you fix $j\ne 1$, the proof is analogous to this one.
 From (\ref{neodndiffur}) it is easily seen that unit map does not move points in projective space at all, because for unit map $A_i(x)x_j-x_iA_j(x)=0$ and therefore $\frac{\p \xi_i}{\p t'}=0$.
%Thus, we have got one more interpretation of the phenomena, when every vector is eigenvector. In this case simply the right-hand side of \ref{neodndiffur} is identically zero. 
The system (\ref{neodndiffur}) can be simplified by choosing $\mu(x)=(x_1)^{s-1}$. Then $\mu(\xi)\equiv1$ and the system becomes:
\be
\frac{\p \xi_i}{\p t'}=A_i(\xi)-A_1(\xi)\xi_i \label{neodndiffurone}
\ee
\subsubsection{$n=2$: the equations are reduced to quadratures}
In the case $n=2$, there is only one non-homogeneous variable: $\xi\equiv\frac{x_2}{x_1}$. The system (\ref{neodndiffurone}) becomes:
\be
 \frac{\p \xi}{\p t'}=A_2(\xi)-A_1(\xi)\xi=P(\xi),
\ee
where $P(\xi)$ is some polynomial of $\xi$ of degree $s+1$ (or lower, if some coefficients of $A$ vanish). This equation is then solved by separation of variables:
\be
 \int\frac{d\xi}{P(\xi)}=t'+C
\ee
When $\xi(t')$ is known, one can find $x_1(t')$ by solving:
\be 
\frac{\p x_1}{\p t'}=\frac{A_1(x)}{\mu(x)}=\frac{(x_1)^sA_1(\xi)}{(x_1)^{s-1}}=x_1 A_1(\xi) \\
\int \frac{dx_1}{x_1}=\int\frac{dt'}{A_1(\xi(t'))}
\ee
$x_1(t')$ can be found from this formula. Then: 
\be
dt'=(x_1(t'))^{s-1}dt \nonumber \\
t+C=\int\frac{dt'}{(x_1(t'))^{s-1}} \nonumber
\ee
From these integrals $t'(t)$ can be found, and, recalling that $x_2=\xi x_1$, one can return to initial variables $t,x_1(t),x_2(t)$.
%From this formula we can find $x_1(t')$, then find $t'(t): dt'=(x_1)^{s-1}dt$, and entirely return to the variables $x_1,x_2=x_1\xi,t$. So we can use this case to test some statement and predictions of general theory. 
This case can be used to test some statement and predictions of general theory. 
\subsection{The "eigenvector" solution \label{eigdiffur}}
As it was shown in previous subsection, if $x_i(t)$ is a solution of 
\be
 \dot{x}_i=A_i^{i_1i_2\cdots i_s}x_{i_1}x_{i_2}x_{i_3}\cdots x_{i_s} 
\label{diffurthree}
\ee
%and simultaneously $x^{(0)}_i$ (initial condition) is
with initial condition $x^{(0)}_i$,
 proportional to an eigenvector $e_i$ of $A$, then $x_i(t)$ evolves so, that
\be
 \frac{x_i(t)}{x_1(t)}=const
\ee
This means, that the solution has a form $x_i(t)=g(t)x^{(0)}_i=f(t)e_i$. If $e_i$ is zero eigenvector, then $x_i=x^{(0)}_i$ and does not change with time. If $e_i$ is unitary eigenvector, %(why there isn't any $\lambda$ in this formula, see \ref{unitaryeig})
 then, substituting $x_i(t)=f(t)e_i$ in (\ref{diffurthree}), one obtains the equation on $f(t)$ with initial condition. In linear and non-linear case the equations are different. In linear case:
\be 
  \frac{d f(t)}{dt}=\lambda_i f(t) \nonumber \\
  f(t)_{t=0}=f_0 \nonumber
\ee
Here $\lambda_i$ is eigenvalue, which corresponds to this eigenvector. Solution of this equation:
\be
f(t)=f_0e^{\lambda_it} \nonumber
\ee
In non-linear case: 
\be
 \frac{d f(t)}{dt}=f(t)^s \\
  f(t)_{t=0}=f_0
\ee
%This equation has different solutions in linear and non-linear cases. 
%\footnote{f(t) in the linear case is not given by this formula, in the linear case $f(t)=f_0e^{\lambda_it}$, where $\lambda_i$ is eigenvalue, which correponds to this eigenvector. This difference is due to the fact, that $\int\frac{dx}{x^s}$ has different formulas in the case $s=1$ and in the case $s\ne 1$}
%In non-linear case:
Solution:
\be
 f(t)=\frac{1}{\left(\frac{1}{f_0^{s-1}}-(s-1)t\right)^{\left(\frac{1}{s-1}\right)}}=\frac{f_0}{(1-(s-1)(f_0)^{s-1}t)^{\left(\frac{1}{s-1}\right)}} \label{reshf} 
\ee
From this formula it is seen, that in non-linear case each solution has a singularity at the moment of time $t_0=\frac{1}{(s-1)f_0^{s-1}}$, i. e. the solution grows infinitely during a finite time. Any stationary point, where the linear term vanishes (see \ref{nonlindiffur}), is unstable in its complex vicinity. To show this it is sufficient to take as initial condition such a vector, that is proportional to unitary eigenvector and $f_0>0$. The vector of initial condition can be made arbitrary small. Then the solution with this initial condition will go at infinity at some finite moment of time. If the map, corresponding to an equation, has a real eigenvector, then this point is unstable in mentioned above sense in its real vicinity.
%\subsubsection{??? When and how a solution of the equation \ref{diffurthree} can grow to infinity at finite time???? }
\subsection{Non-linear condition of instability \label{uslstab} }
Any stable point, in which linear term equals zero (see sect.\ref{nonlindiffur}) and main non-linear term has unitary eigenvector, is unstable in its complex vicinity. To show this, it is sufficient to take as initial condition a vector, which is proportional to a unitary eigenvector and $f_0>0$. This vector can be done arbitrarily small. Then the solution will go at infinity in a some finite time. If the map has real unitary eigenvector, this point is unstable in its real vicinity. The question of stability of the point of equilibrium of the map without unitary eigenvectors is open. This case is encountered rarely. The map should be very degenerate, if it has no unitary eigenvectors. For example, the map 
\be
\left(
\ba{c}
x_1 \\
x_2 
\ea
\right) \rightarrow \left(
\ba{c}
0 \\
ax_1x_2
\ea
\right) \nonumber
\ee
has no unitary eigenvectors, and two zero eigenvectors $(0,1)$ and $(1,0)$. This is the result of solving of equations (\ref{zeroeig}) and (\ref{oneeig}). This map has only one non-zero component of six possible. So, here is a brief summary:
\begin{enumerate}
\item If a map has an unitary eigenvector, the origin is unstable.
\item If a map has non-zero resultant, the origin is unstable
\item If a map has no unitary eigenvectors, one cannot make any definite prediction about stability/instability of the origin.
%one needs investigate the stability/instability in every particular case.
\end{enumerate}
If the resultant of the map does not equal to $0$, the map cannot have zero eigenvectors. So, it has unitary eigenvectors and one comes to item 1.
\subsection{The example of equations with vanishing linear term \label{examplevan}}
Consider the system of differential equations of chemical kinetics of this system of reactions and reagents:
\be
 %C_6H_{12}+H_2O \leftrightarrow C_2H_6O+C_4H_8 \\
 %C_2H_6O+C_6H_{12}\leftrightarrow H_2O+2 C_4H_8 \\
 C+D \rightarrow A+B \nonumber  \\
 C+A \leftrightarrow B+D+D \nonumber
\ee
%The reader should be aware that this is very strange example from the point of view of chemistry, because the reactions either do not occur, either occur but with many additional ways of reaction. 
%Consider this idealized example. 
$A,B,C,D$ represent the reagents. $X_1,X_2,X_3,X_4$ are, respectively, partial concentrations of $A,B,C,D$. The condition of mass conservation in this system looks like this\footnote{The mass conservation law is not always so simple. It depends on the particular substances. It can occur that there are several independent equations of mass conservation, for each element. Very simplified approximation is considered here.}:
\be
X_1+X_2+X_3+X_4=1 \label{massconserv}
\ee
The equations of chemical kinetics for this system:
\bs{l}
\dot{X}_1=K_{34}X_3X_4-K_{31}X_1X_3+K_{24}X_2(X_4)^2 \\
\dot{X}_2=K_{34}X_3X_4+K_{31}X_1X_3-K_{24}X_2(X_4)^2  \\
\dot{X}_3=-K_{34}X_3X_4-K_{31}X_1X_3+K_{24}X_2(X_4)^2  \\
\dot{X}_4=-K_{34}X_3X_4+2K_{31}X_1X_3-2K_{24}X_2(X_4)^2,  \label{chemdiffur}
\es
where dot denotes time derivative, $K_{34},K_{31},K_{24}$ are rate coefficients of the reactions. Actually, they depend on the temperature and pressure. If these dependencies are neglected, the equations of type (\ref{diffurtwo}) are obtained.
From (\ref{massconserv}) $X_4$ can be expressed through $X_1,X_2,X_3$, and there will be only three variables $X_1,X_2,X_3$. One of stationary points of this system is:
\bs{l}
X_1^{(0)}=0 \label{statpoint} \\
X_2^{(0)}=1  \\
X_3^{(0)}=0 
\es
Denote now $\delta_1=X_1-X_1^{(0)},\delta_2=X_2-X_2^{(0)},\delta_3=X_3-X_3^{(0)}$. The linear term in the expansion of right-hand sides of (\ref{chemdiffur}) vanishes near the point (\ref{statpoint}). In the new variables, $\delta_1,\delta_2,\delta_3$, the system (\ref{chemdiffur}) is:
\be
\dot{\delta}_1=K_{24}\de_1^2+K_{24}\de_2^2+(K_{24}-K_{34})\de_3^2+2K_{24}\de_1\de_2+(2K_{24}-K_{34})\de_2\de_3+(2K_{24}-K_{34}-K_{31})\de_1\de_3+ \nonumber \\
+{\rm cubic\quad terms \quad} \label{deltaone} \\
\dot{\delta}_2=-K_{24}\de_1^2-K_{24}\de_2^2-(K_{24}+K_{34})\de_3^2-2K_{24}\de_1\de_2+(K_{31}-K_{34}-2K_{24})\de_3\de_1-(2K_{24}+K_{34})\de_2\de_3+ \nonumber \\
+{\rm cubic\quad terms \quad} \label{deltatwo} \\
\dot{\delta}_3=K_{24}\de_1^2+2K_{24}\de_1\de_2+(K_{34}+2K_{24}-K_{31})\de_3\de_1+K_{24}\de_2^2+(2K_{24}+K_{34})\de_2\de_3+(K_{24}+K_{34})\de_3^2+ \nonumber \\
+{\rm cubic\quad terms\quad} \label{deltathree}
\ee
To analyse the stability of stationary point (\ref{statpoint}), one should consider only the term in expansion of the lowest degree, in this case, of second degree. After neglecting cubical terms in (\ref{deltaone})-(\ref{deltathree}), one obtains the  differential equation with homogeneous quadratic right-hand side, i. e. the equation of type (\ref{diffurtwo}).
\subsubsection{Investigation of stability of this example}
To know whether the origin of the system of equations (\ref{deltaone}-\ref{deltathree}) is stable or not, one should consider the system of equations:
\be
\left(\ba{c} 
\delta_1 \\
\delta_2 \\
\delta_3
\ea\right)\mapsto
\left(\ba{c}
K_{24}\de_1^2+K_{24}\de_2^2+(K_{24}-K_{34})\de_3^2+2K_{24}\de_1\de_2+(2K_{24}-K_{34})\de_2\de_3+(2K_{24}-K_{34}-K_{31})\de_1\de_3  \\
-K_{24}\de_1^2-K_{24}\de_2^2-(K_{24}+K_{34})\de_3^2-2K_{24}\de_1\de_2+(K_{31}-K_{34}-2K_{24})\de_3\de_1-(2K_{24}+K_{34})\de_2\de_3  \\
K_{24}\de_1^2+2K_{24}\de_1\de_2+(K_{34}+2K_{24}-K_{31})\de_3\de_1+K_{24}\de_2^2+(2K_{24}+K_{34})\de_2\de_3+(K_{24}+K_{34})\de_3^2 
\ea\right)
\ee
All unitary eigenvectors satisfy the equation:
\be
\de_1= K_{24}\de_1^2+K_{24}\de_2^2+(K_{24}-K_{34})\de_3^2+2K_{24}\de_1\de_2+(2K_{24}-K_{34})\de_2\de_3+\sls
 +(2K_{24}-K_{34}-K_{31})\de_1\de_3 \sls
\de_2= -K_{24}\de_1^2-K_{24}\de_2^2-(K_{24}+K_{34})\de_3^2-2K_{24}\de_1\de_2+(K_{31}-K_{34}-2K_{24})\de_3\de_1-\sls
(2K_{24}+K_{34})\de_2\de_3 \sls
\de_3=K_{24}\de_1^2+2K_{24}\de_1\de_2+(K_{34}+2K_{24}-K_{31})\de_3\de_1+K_{24}\de_2^2+(2K_{24}+K_{34})\de_2\de_3+\sls
+(K_{24}+K_{34})\de_3^2 
\ee
One can notice, that $f_2+f_3=0$, so the resultant of this map equals 0. So one cannot use the second item from sect.\ref{uslstab}. $f_2+f_3=0$, and therefore, $\de_3=-\de_2$. Substitute it in first two equations. Then one can exclude $\de_1$ from the equations and obtain the equation with single variable $\de_2$:
\be
 2(K_{34}^2+K_{31}K_{34})\de_2^2+(K_{24}-3K_{34}-K_{31})\de_2+1=0
\ee
This equation always has roots, so the point $(0,1,0)$ is unstable in its complex vicinity, since there are complex unitary eigenvectors. It has real roots, when $(K_{24}-3K_{34}-K_{31})^2-8 (K_{34}^2+K_31K_34)>0$, and then it is unstable in its real vicinity.
% We do not know the answer 
\section{Example of finding degeneracies and peculiarities of non-linear map \label{example}  }
In this section an example of quadratic map of two variables in canonical form is considered. Nevertheless, {\it all} results, formulated in sections \ref{apriorsovp}, \ref{apriores}, \ref{aprioreq}, can be obtained for any non-linear map in any coordinates. The only difference (between this example and general situation) is that the developed tools are sufficient to {\it fully} investigate and classify this example, but in general situation higher resultants, higher complanarts, etc can be needed. %But the information, obtained in this section for our example, can be obtained for any map in any coordinates (we will discuss how to do the same things in general situations).
Only the case when two unitary eigenvectors are chosen to be basis vectors is considered. \\
%, other cases are similar (and which are the other cases, see sect. \ref{excanrepres}).  \\
Our example is:
\be
\ba{ccc}
 x_1 & \rightarrow & x_1^2+2ax_1x_2 \\
 x_2 & \rightarrow & x_2^2+2bx_1x_2 \label{systexample}
\ea
\ee
$a$ and $b$ are the parameters, defining a map instead of eigenvalues (see \ref{canrepres}). 
\subsection{General analysis \label{exampleig}}
By construction of canonical form of the map, (\ref{systexample}) has two unitary eigenvectors, namely $e^{(1)}\equiv(1,0)$ and $e^{(2)}\equiv(0,1)$. They are unitary eigenvectors under any values of $a$ and $b$. In general case this map has $c_{n|s}=\frac{s^n-1}{s-1}=3$ ($s=2,n=2$) eigenvectors. 
% We consider the case when $R\ne 0$, 
Here are the equations (\ref{oneeig},\ref{zeroeig},\ref{oneeigodn}) in this case.
\bs{c}
 x_1^2+2ax_1x_2=x_1 \label{oneeigour}\\
 x_2^2+2bx_1x_2=x_2 %\label{oneeigourtwo}
\es
This is a system (\ref{oneeig}) in this case. One should solve this system of equations to find unitary eigenvectors.
% and we will seek for unitary eigenvectors by solving these equations.
\bs{c}
 x_1^2+2ax_1x_2=0 \label{zeroeigour} \\
 x_2^2+2bx_1x_2=0 % \label{zeroeigourtwo}
\es
This is a system (\ref{zeroeig}) in this case. One should solve this system of equations to find zero eigenvectors.
%, and we will seek for zero eigenvectors by solving these equations.
\bs{c}
 x_1^2+2ax_1x_2=x_1y \\
 x_2^2+2bx_1x_2=x_2y, \label{oneeigodnour}
\es
This is a system (\ref{oneeigodn}) in this case, and all eigenvectors (zero, non-zero unitary) are its solutions. Additional homogenizing variable is also called $y$ here, as in (\ref{oneeigodn}).
%where $y\equiv x_3$ - is analogue to $x_{n+1}$ in (\ref{oneeigodn}).
%\footnote{Appropriate $y$ value for non-zero eigenvector is any $y\ne 0$, appropriate $y$ for unitary eigenvector is 1 or any complex root of 1 of degree $s-1$, and appropriate $y$ value for zero eigenvector is 0}
A solution of (\ref{oneeigodnour}), $(x_1,x_2,y)$ with $y=0$ corresponds to zero eigenvector. This zero eigenvector is $(x_1,x_2)$. A solution of (\ref{oneeigodnour}), $(x_1,x_2,y)$ with $y\ne 0$ corresponds to non-zero eigenvector. This non-zero eigenvector is $(x_1,x_2)$. Since the system (\ref{oneeigodnour}) is a system of homogeneous equations, the solutions of this system can be scaled by any non-zero number. To get unitary eigenvector from a non-zero eigenvector, one should rescale a solution of (\ref{oneeigodnour}) with $y\ne 0$ by dividing it by $y$: then a solution of (\ref{oneeigodnour}) becomes $(x_1/y,x_2/y,1)$ and unitary eigenvector is $(x_1/y,x_2/y)$. 
\subsubsection{Preliminary information about eigenvectors from complanart \label{apriorsovp}}
 (\ref{oneeigodnour}) is a system of two homogeneous equations of three variables. Such system possesses a complanart. Complanart of this system is calculated in sect \ref{sovpab}, and it equals
\be 
 C=(1-2a)^4(1-2b)^4 \nonumber
\ee 
When it vanishes,
% that is when $a=1/2$ or $b=1/2$, 
 the system (\ref{oneeigodnour}) has three complanar in the space $x_1,x_2,y$ roots. A system (\ref{oneeigodnour}) is a system of two quadratic equations, so in general case it has $2\cdot2=4$ solutions. Two solutions of (\ref{oneeigodnour}) correspond to known unitary eigenvectors, $e^{(1)}\equiv(1,0)$ and $e^{(2)}\equiv(0,1)$. These solutions of (\ref{oneeigodnour}) are $\Lambda^{(1)}\equiv(1,0,1)$ and $\Lambda^{(2)}\equiv(0,1,1)$. There should be one more eigenvector in this case, since the map (\ref{systexample}) in general case has $c_{n|s}=c_{2|2}=3$ eigenvectors. Denote this eigenvector by $e^{(3)}\equiv(e^{(3)}_1,e^{(3)}_2)$. Then the solution of (\ref{oneeigodnour}), corresponding to this eigenvector, is  $\Lambda^{(3)}\equiv(e^{(3)}_1,e^{(3)}_2,e^{(3)}_3)$. If $e^{(3)}$ is zero eigenvector, $e^{(3)}_3=0$, if $e^{(3)}$ is non-zero eigenvector, $e^{(3)}_3\ne0$, and $e^{(3)}_3=1$ if $e^{(3)}$ is unitary eigenvector. And the last solution of (\ref{oneeigodnour}) is $\Lambda^{(4)}\equiv(0,0,1)$, this is a solution (\ref{neint}) for this system. Vanishing of complanart means that $\eps^{i_1i_2i_3}\Lambda_{i_1}^{(j_1)}\Lambda_{i_2}^{(j_2)}\Lambda_{i_3}^{(j_3)}=0$ with at least one triple of indices $j_1,j_2,j_3, \quad j_1\ne j_2,j_1\ne j_3,j_2\ne j_3$. Particularly, vanishing of complanart means vanishing of at least one of the following expressions:
\be
\eps^{i_1i_2i_3}\Lambda_{i_1}^{(1)}\Lambda_{i_2}^{(2)}\Lambda_{i_3}^{(3)}=e^{(3)}_1-e^{(3)}_2-e^{(3)}_3 \label{onetwothree} \\
\eps^{i_1i_2i_3}\Lambda_{i_1}^{(1)}\Lambda_{i_2}^{(2)}\Lambda_{i_3}^{(4)}=1 \label{onetwofour} \\
\eps^{i_1i_2i_3}\Lambda_{i_1}^{(2)}\Lambda_{i_2}^{(3)}\Lambda_{i_3}^{(4)}=e^{(3)}_1 \label{twothreefour} \\
\eps^{i_1i_2i_3}\Lambda_{i_1}^{(1)}\Lambda_{i_2}^{(3)}\Lambda_{i_3}^{(4)}=e^{(3)}_2 \label{onethreefour}
\ee
The solution $\Lambda^{(4)}$ enters in (\ref{onetwofour}),(\ref{twothreefour}), (\ref{onethreefour}), so these equations are nothing but conditions of pairs of eigenvectors to be complanar, see discussion in sect \ref{sovpappl}. The equation (\ref{onetwofour}) controls complanarity of $e^{(1)}$ and $e^{(2)}$, (\ref{twothreefour}) - of $e^{(2)}$ and $e^{(3)}$, and (\ref{onethreefour}) - of $e^{(1)}$ and $e^{(3)}$. The condition (\ref{onetwothree}) has no such a simple interpretation. So, if $a=1/2$ or $b=1/2$, complanart vanishes, and one expects that either $e^{(2)}$ and $e^{(3)}$, or $e^{(1)}$ and $e^{(3)}$ will be collinear, or the condition (\ref{onetwothree}) will hold.
\subsubsection{Preliminary information about eigenvectors from resultant \label{apriores} }
Resultant of the map (\ref{systexample}) equals:
\be
 R\{A\}=1-4ab \nonumber
\ee
As it was stated in sect \ref{numeig}, zero eigenvectors exist iff the resultant of the map equals $0$. So one expects, that if $1-4ab=0$, there will be zero eigenvectors, and if $1-4ab\ne 0$, there will be no zero eigenvectors, and will be only non-zero eigenvectors.
\subsubsection{Preliminary information about eigenvectors from the equation $x_iA_j-x_jA_i$  \label{aprioreq}}
As it was stated in sect \ref{unitmaps}, if polynomials $x_iA_j(x)-x_jA_i(x)$ are identically zero, the map $A$ is unit map. For the map (\ref{systexample}), there is only one equation:
\be
 x_1A_2(x)-x_2A_1(x)\equiv x_1(x_2^2+2bx_1x_2)-x_2(x_1^2+2ax_1x_2)\equiv x_1x_2^2(1-2a)-x_1^2y(1-2b)=0
\ee
So, if $a=1/2$ and $b=1/2$, the map $A$ will be unit map, every vector of the space will be eigenvector of $A$, every polynomial of degree $s-1\quad \mu(x)$ will be an eigenvalue of the map $A$, and characteristic polynomial of $A$ will be identically zero. 
\subsection{Eigenvectors}
%Now solve this system.
Besides two solutions $(x_1,x_2,y)=(0,1,1)$ and $(x_1,x_2)=(1,0,1)$, the system (\ref{oneeigodnour}) has the following solution:
% In homogeneous variables the solution () is:
\be
\left(
\ba{c}
x_1 \\
x_2 \\
y
\ea
\right)=\left(
\ba{c}
1-2a \\
1-2b \\
1-4ab
\ea
\right)
\label{neodnsolution}
\ee
%It should be determined, whether the solution (\ref{neodnsolution}) corresponds to zero or non
%The solutions of (\ref{oneeigodnour}) 
%After we have solved homogeneous system, we should convert its solutions to eigenvectors, i. e. to the solutions of (\ref{oneeigour}) or (\ref{zeroeigour}). 
If $y$ value
%\footnote{It is not the random coincidence - $y$ value is equal to resultant of a map}
 is equal to zero, (\ref{neodnsolution}) corresponds to zero eigenvector:
\be
v^{(1)}\equiv\left(
\ba{c}
1-2a \\
1-2b
\ea
\right)
\ee
(of course, the normalization is arbitrary). In this case $e^{(3)}_3=0$. \\
If $y$ value is not equal to zero, (\ref{neodnsolution}) corresponds to non-zero eigenvector. In this case, the solution vector can be renormalized:
\be
\left(
\ba{c}
\frac{1-2a}{1-4ab} \\
\frac{1-2b}{1-4ab} \\
 1
\ea
\right)
\ee
to get unitary eigenvector:
\be
e^{(3)}\equiv
\left(
\ba{c}
\frac{1-2a}{1-4ab} \\
\frac{1-2b}{1-4ab}
\ea
\right) \label{unitsolution}
\ee
In this case $e^{(3)}_3=1$. In agreement with predictions of sect \ref{apriores}, zero eigenvector exists iff $1-4ab=0$, or when the resultant vanishes. In agreement with predictions of sect \ref{apriorsovp}, if $a=1/2$ or $b=1/2$, there are two coinciding eigenvectors. If $a=1/2$, (\ref{unitsolution}) coincides with $e^{(2)}$, and if $b=1/2$, (\ref{unitsolution}) coincides with $e^{(1)}$.
%With agreement with the predictions made by us with help of complanart, if $a=1/2$ or $b=1/2$, the system has only two eigenvectors, but one of them is of double multiplicity.
%As you can see from the last formula, the case of $a=1/2$ {\it and} $b=1/2$ is some special case, and we will consider this case later. 
The prediction of sect \ref{aprioreq}, namely, the case $a=1/2$ {\it and} $b=1/2$ requires separate consideration. All the components of (\ref{neodnsolution}) are now zero. The equation on eigenvectors/eigenvalues with $a=1/2, b=1/2$:
\bs{c}
 x_1^2+x_1x_2=x_1y \\
 x_2^2+x_1x_2=x_2y \label{systunit}
\es
The case of unit map is the case of decreasing range of the system (\ref{oneeigodnour}). Unfortunately, there have been developed no clear methods for determining whether the system of non-linear equations is degenerate. Complanart is an attempt to full this gap, but it is easily seen from this example that it does not distinguish the cases when two vectors coincide and the decreasing of range of a system.
% In higher dimensions, it is expected that there are the maps with decreased range of the system (\ref{}), but which are not obligatory unit maps.
 Only eigenvectors, which are not equal to known $(0,1)$ and $(1,0)$, are interested in now. So, first equation of (\ref{systunit}) can be divided by $x_1$ and the second one can be divided by $x_2$. 
%Since we want to find eigenvectors, which are not equal to known $(0,1)$ and $(1,0)$, we can divide first equation by $x_1$ and second by $x_2$:
\bs{c}
x_1+x_2=y \\
x_1+x_2=y \nonumber
\es
The equation $x_1+x_2=y$ has projective one-dimensional space of solutions: one can set $y=1$ (to obtain unitary eigenvectors), $x_2=C$ and $x_1$ will be $1-C$. For any $C$ the vector
\be
\left(
\ba{c}
C \\
1-C
\ea
\right)
\ee
is unitary eigenvector. But there is also zero eigenvector, $(x_1,x_2,y)=(-1,1,0)$. Arbitrary vector of the space $(x_1,x_2)$ is zero eigenvector, if $x_1+x_2=0$, and $(x_1,x_2)$ is non-zero eigenvector, if $x_1+x_2\ne 0$. 
% Let's look again on this classification from the point of view of characteristical polynomial. Let's write
\subsection{Eigenvalues}
Polynomial $\lambda$, corresponding to unitary eigenvector $e^{(i)}$ is found from equation $\lambda(e^{(i)})=1$, see sect \ref{unitaryeig}. This equation for $e^{(1)}$ is $\lambda_1=1$, i. e. to the eigenvector $e^{(1)}$ corresponds the line $\lambda_1=1$ on the plane $\lambda_1,\lambda_2$. And for $e^{(2)}$: $\lambda_2=1$, i. e. to the eigenvector $e^{(2)}$ corresponds the line $\lambda_2=1$ on the plane $\lambda_1,\lambda_2$. For $e^{(3)}$: $\lambda_1\frac{1-2a}{1-4ab}+\frac{1-2b}{1-4ab}\lambda_2=1$, or
\be
 \lambda_1(1-2a)+\lambda_2(1-2b)=1-4ab \label{lambdaethree}
\ee
This is also a line at $\lambda_1,\lambda_2$ plane.
%it is necessary to be emphasized here, that to one eigenvector corresponds not one $\lambda$, but the entire plane in the space of all polynomials of degree $s-1$ of n variables, and this plane is defined by $\lambda(e^{(i)})=1$ or $\lambda(v^{(i)})=0$.
 To find $\lambda$, corresponding to zero eigenvector, one should solve $\lambda(v^{(i)})=0$, i. e. in this case $\lambda_1(1-2a)+\lambda_2(1-2b)=0$. But (\ref{lambdaethree}) is equivalent to this equation, because the resultant, namely $1-4ab$, turns to zero if there is a zero eigenvector. It is not a random coincidence, namely the term $\lambda_1(1-2a)+\lambda_2(1-2b)-(1-4ab)$ is a factor in the decomposition of characteristic polynomial. It is easily seen, that in the case of $a=1/2$ and $b=1/2$ (or when our map is unit map), {\it any} $\lambda$ is eigenvalue.
\subsection{Characteristic polynomial}
The system of equations (\ref{eigvect}) for the map (\ref{systexample}):
\bs{c}
 x_1^2+2ax_1x_2=(\lambda_1x_1+\lambda_2x_2)x_1 \\
 x_2^2+2bx_1x_2=(\lambda_1x_1+\lambda_2x_2)x_2 \nonumber
\es
Characteristic polynomial is the resultant of the system:
\bs{c}
 x_1^2+2ax_1x_2-(\lambda_1x_1+\lambda_2x_2)x_1=0 \\
 x_2^2+2bx_1x_2-(\lambda_1x_1+\lambda_2x_2)x_2=0 \nonumber
\es
The characteristic polynomial is equal:
\be
Ch_A(\lambda)=(1-\lambda_1)(1-\lambda_2)(1-4ab-\lambda_1(1-2a)-\lambda_2(1-2b))
\ee
Here we would like to emphasize one more time: decomposability of characteristic polynomial is {\it non-trivial} property. The polynomial $1+(\lambda_1)^2+(\lambda_2)^2$, for example, cannot be decomposed on linear on $\lambda_1,\lambda_2$ factors. If the resultant does not equal to 0:
\be
 Ch_A(\lambda)=(1-4ab)(1-\lambda(e^{(1)}))(1-\lambda(e^{(2)}))(1-\lambda(e^{(3)})) \nonumber
\ee
in full agreement with (\ref{razlnondegchar}). If resultant of a map is equal to zero:
\be
 Ch_A(\lambda)=(1-\lambda(e^{(1)}))(1-\lambda(e^{(2)}))\lambda(v^{(1)}) \nonumber
\ee
in full agreement with (\ref{razlchar}). If $a=1/2$ and $b=1/2$ (i. e. unit map), characteristic polynomial is identically zero, in full agreement with s.\ref{unitmaps}.
%The "strange" phenomena when there is a continuum of eigenvectors (when $a=1/2$ and $b=1/2$) is nothing but turning the characteristic polynomial to zero (the factor $(1-4ab-\lambda_1(1-2a)-\lambda_2(1-2b))$ is identically zero when $a=1/2$ and $b=1/2$). 
%As you can see from the last formula, the case of $a=1/2$ {\it and} $b=1/2$ is some special case, and we will consider this case later. 

\subsection{Phase diagram}
Differential equation, which corresponds to the map (\ref{systexample})  , is:
\bs{c}
\dot{x_1}=x_1^2+2ax_1x_2 \\
\dot{x_2}=x_2^2+2bx_1x_2
\es
%\subsubsection{The phase portrait}
%Now we classify all possible cases of eigenvectors/eigenvalues of this map and all possible phase portraits.
\subsubsection[$a\ne 1/2,b\ne 1/2,1-4ab\ne 0$]{$\mathbf{a\ne 1/2,b\ne 1/2,1-4ab\ne 0}$}
\begin{figure}[h!]
\begin{center}
\includegraphics[width=0.5\linewidth]{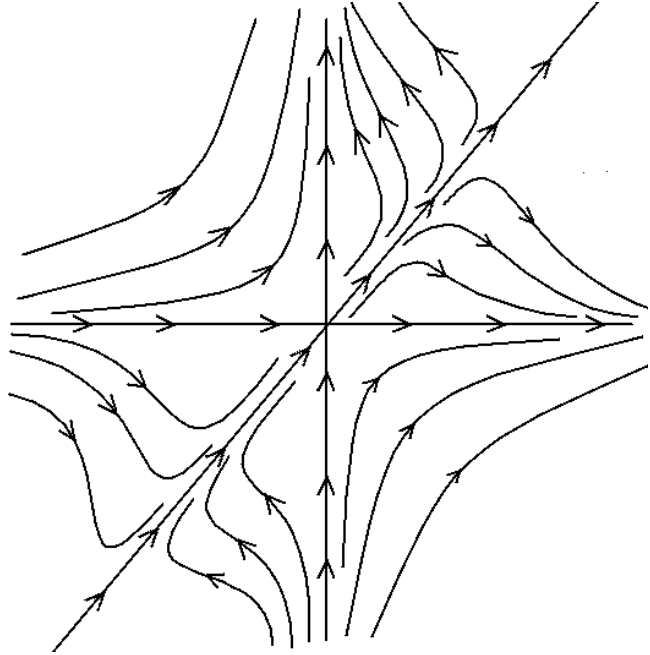}
\caption{The phase portrait with $a=b=-1$. There are no degeneracies and peculiarities, there are three unitary eigenvectors:$(0,1),(1,0),(3/5,3/5)$}
\label{gencase}
\end{center}
\end{figure}

This is the case of absence of any degeneracies and pecularities. There are three unitary eigenvectors, $(1,0);(0,1);(\frac{1-2a}{1-4ab},\frac{1-2b}{1-4ab})$. Fig.\ref{gencase} is the phase portrait of this system with $a=-1,b=-1$.

\subsubsection[$a\ne 1/2,b\ne 1/2,1-4ab=0 $]{$\mathbf{a\ne 1/2,b\ne 1/2,1-4ab=0}$}
\begin{figure}[h!]
\begin{center}
\includegraphics[width=0.5\linewidth]{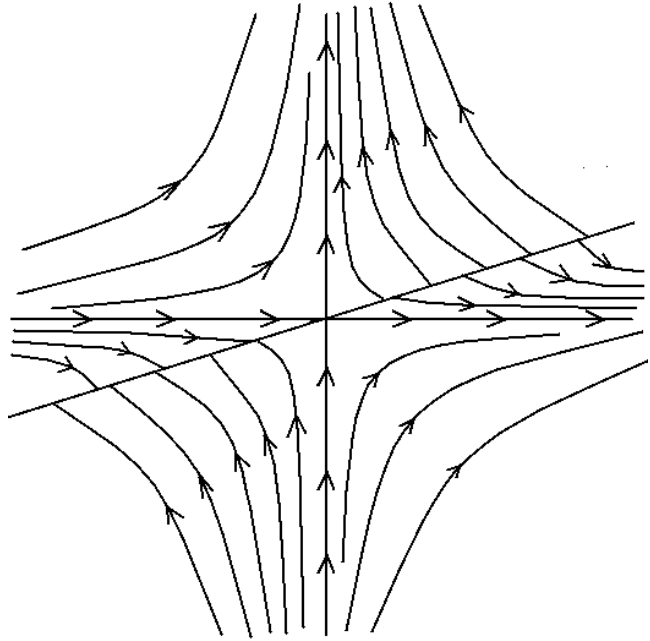}
\caption{The phase portrait with $a=-1,b=-1/4$. The map is degenerate, so it has zero eigenvector $(3,2)$. All the points of the line $\frac{x}{3}=\frac{y}{2}$ are stationary points.}
\label{nullsv}
\end{center}
\end{figure}

The map in this case is degenerate, i. e. it has a zero eigenvector $(1-2a,1-2b)$ with full accordance with sect \ref{apriores}. Each point, which lie on the line $\frac{x_1}{1-2a}=\frac{x_2}{1-2b}$ (this is a line of zero eigenvectors), is a stationary point. Fig.\ref{nullsv} is the phase portrait of this system with $a=-1,b=-1/4$.

\subsubsection[$a\ne 1/2,b=1/2$]{$\mathbf{a\ne 1/2,b=1/2}$}
\begin{figure}[h!]
\begin{center}
\includegraphics[width=0.5\linewidth]{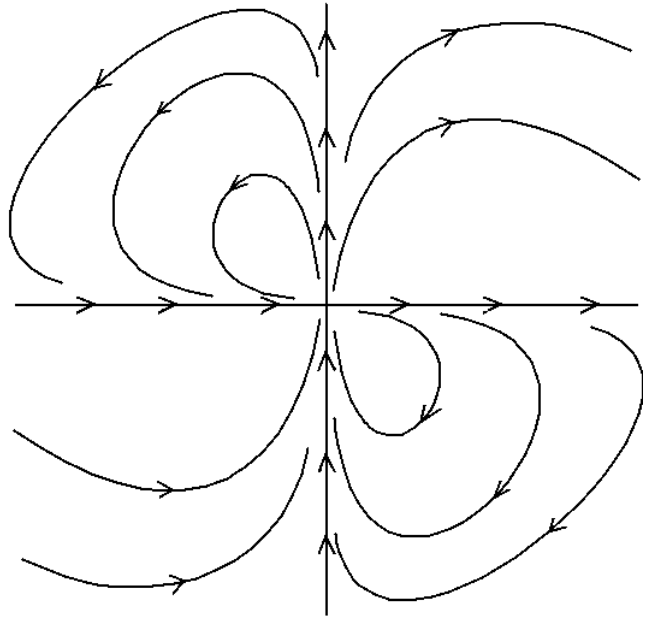}
\caption{The phase portrait with $a=1,b=1/2$. Because the complanart of the map equals 0, there is double eigenvector $(1,0)$. Trajectories at upper right corner tend to eigenvector $(1,0)$ in projective space and at lower right corner trajectories move out of eigenvector $(1,0)$ in projective space. This is a general property of double eigenvectors.}
\label{sovpsv}
\end{center}
\end{figure}

Because the complanart of this map equals $0$, there is a double eigenvector of this map, with full accordance with sect \ref{apriorsovp}. In this case the map has only two unitary eigenvectors, $(0,1)$ is "simple" eigenvector (with multiplicity 1), and $(1,0)$ with multiplicity 2. The eigenvectors with multiplicity 2 have a special property: the phase trajectories tend in projective space to this eigenvector from one side, and they tend out of it in projective space at other side of this eigenvector. On fig.\ref{sovpsv} at upper right corner trajectories tend to eigenvector $(1,0)$ in projective space (namely, $x_1\rightarrow \infty,x_2\rightarrow\infty,\frac{x_2}{x_1}\rightarrow 0$) and at lower right corner trajectories move out of eigenvector $(1,0)$ in projective space. The equation on $x_2/x_1 \equiv \zeta$ in the vicinity of $0$ looks like 
\be
\dot{\zeta}=C\zeta^2,
\ee
 where $C$ is a constant, depending on $a$ and $b$. This projective equation explains such a behaviour: $\dot\zeta$ has the same signs at both sides from stationary point, so the solution approaches from one side and move away from other side. Near a double eigenvectors, the linear term in projective equation always vanishes.
% Near an eigenvector with multiplicity $2$ linear term always vanishes and it is a stationary point in projective space with unusual property: the trajectories approach it from one side and leave at the other side.
  Fig.\ref{sovpsv} is the phase portrait of the system with $a=1,b=1/2$. The case with $a=1/2, b\ne1/2$ reduces to this case by substitution $x_1\rightarrow x_2, x_2\rightarrow x_1$.

\subsubsection[$a=1/2,b=1/2$]{$\mathbf{a=1/2,b=1/2}$}
\begin{figure}[h!]
\begin{center}
\includegraphics[width=0.5\linewidth]{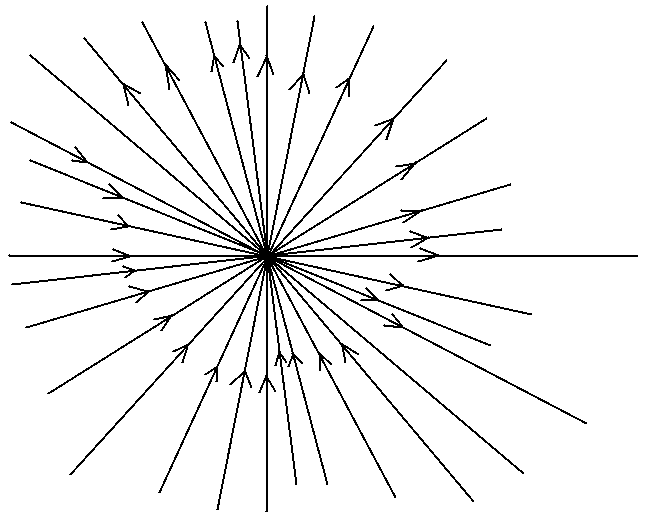}
\caption{The phase portrait with $a=1/2,b=1/2$. Unlike all previous cases, all the phase trajectories are straight lines. This is because this map is unit map, and all vectors of the space are eigenvectors. Vectors, proportional to $(-1,1)$ vector, are zero eigenvectors, and all other vectors are non-zero eigenvectors.}
\label{pictunitmap}
\end{center}
\end{figure}

According to predictions of \ref{aprioreq}, this map is unit map and any vector is an eigenvector. Fig.\ref{pictunitmap} is the phase portrait in this case.
\subsubsection{Phase diagram without unitary eigenvectors}
\begin{figure}[h!]
\begin{center}
\includegraphics[width=0.5\linewidth]{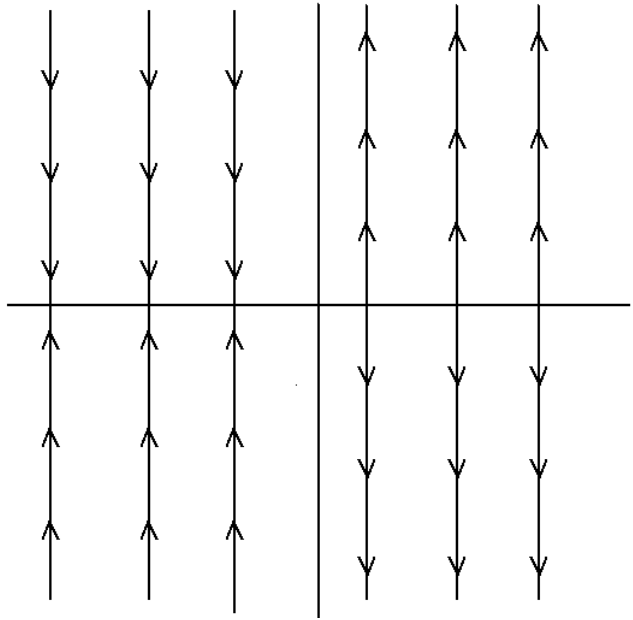}
\caption{Phase diagram of (\ref{systexamplevur}). $Oy$ axe is vertical line, which has no arrows. The origin is unstable in this case.}
\label{pictdvanull}
\end{center}
\end{figure}

The example of map without unitary eigenvectors is considered here. There are no methods to determine whether the point which has not unitary eigenvectors is stable or not. The example is:
\be
\ba{ccc}
 x_1 & \rightarrow & 0 \\
 x_2 & \rightarrow & bx_1x_2 \label{systexamplevur}
\ea
\ee
The basis consists of two zero eigenvectors. Besides these vectors, there are no other eigenvectors. In this particular case the point $(0,0)$ is unstable. Phase diagram is at the Fig.\ref{pictdvanull}.

\section{Acknowledgements}
We want to thank A. Morozov and Sh. Shakirov for useful discussions. This work is supported by grant for support of scientific schools NSh-3036.2008.2. and by grant of RFBR 09-02-00393.
%07-02-00645.

\end{document}